\documentclass[useAMS,usenatbib]{mn2e}

\usepackage[psamsfonts]{amssymb}
\usepackage[utf8]{inputenc}
\usepackage[T1]{fontenc}
\usepackage[dvips]{graphicx}
\usepackage{amsmath,alltt}
\usepackage{multirow}
\usepackage{rotating}
\usepackage{lscape}

\newcommand{\bdm}{\begin{displaymath}}
\newcommand{\edm}{\end{displaymath}}
\newcommand{\beq}{\begin{equation}}
\newcommand{\eeq}{\end{equation}}
\newcommand{\bit}{\begin{itemize}}
\newcommand{\eit}{\end{itemize}}
\newcommand{\ben}{\begin{enumerate}}
\newcommand{\een}{\end{enumerate}}
\newcommand{\bfi}{\begin{figure}[htb]}
\newcommand{\bpfi}{\begin{figure}[p]}

\title[NSC formation in energy-space]{Nuclear star cluster formation in energy-space}
\author[Leigh N. W. C., Georgiev I. Y., B\"oker T., Knigge C., den Brok M.]{Nathan W. C. Leigh$^{1,2}$, Iskren Y. Georgiev$^{3}$, 
Torsten B\"oker$^{3}$, Christian Knigge$^{4}$, 
\newauthor
Mark den Brok$^{5}$
\thanks{E-mail: nleigh@amnh.org (NL); tboeker@cosmos.esa.int (TB); C.Knigge@soton.ac.uk (CK); iskren.y.g@gmail.com (IG); denbrok@physics.utah.edu}\\
$^{1}$Department of Astrophysics, American Museum of Natural History, Central Park West and 79th Street, New York, NY 10024 \\
$^{2}$Department of Physics, University of Alberta, CCIS 4-183, Edmonton, AB T6G 2E1, Canada \\
$^{3}$European Space Agency, Space Science Department, Keplerlaan 1,
2200 AG Noordwijk, The Netherlands \\
$^{4}$School of Physics and Astronomy, University of Southampton,
Highfield, Southampton, SO17 1BJ, United Kingdom \\
$^{5}$Department of Physics and Astronomy, University of Utah, 115 South 1400 East, Salt Lake City, UT 84112, USA}
\begin{document}

\pagerange{\pageref{firstpage}--\pageref{lastpage}} \pubyear{2011}

\maketitle

\label{firstpage}

\begin{abstract}
In a virialized stellar system, the mean-square velocity is a direct tracer of the 
energy per unit mass of the system.  Here, we exploit this to estimate and compare 
root-mean-square velocities for a large sample of nuclear star clusters and their host 
(late- or early-type) galaxies.  Traditional observables, such as the 
radial surface brightness and second-order velocity moment profiles, are subject to short-term 
variations due to individual episodes of matter infall and/or star formation.  The total mass, 
energy and angular momentum, on the other hand, 
are approximately conserved.  Thus, the total energy and angular momentum more directly probe 
the formation of galaxies and their nuclear star clusters, by offering access to 
more fundamental properties of the nuclear cluster-galaxy system than traditional observables.  
We find that there is a strong correlation, in fact a near equality, between the root-mean-square 
velocity of a nuclear star cluster and that of its host.  Thus, the energy per unit mass of a 
nuclear star cluster is always comparable to that of its host galaxy.  
We interpret this as evidence that nuclear star clusters do not form independently 
of their host galaxies, but rather that their formation and subsequent evolution are coupled.  
We discuss how our results can potentially be used to offer a clear and observationally testable 
prediction to distinguish between the different nuclear star cluster formation scenarios, and/or 
quantify their relative contributions.  

\end{abstract}


\section{Introduction} \label{intro}
Recent studies have revealed that nuclear star clusters (NSCs) occur
commonly in galaxies of every type \citep[e.g.][]{binggeli84,kormendy04}.  
These massive and compact star clusters occupying the centres 
of galaxies are characterized by small effective radii and high
central surface brightnesses that help to distinguish them from the
surrounding bulge and disk.  

Several theories have been proposed to explain the origins of NSCs, but 
two in particular have emerged as leading scenarios.  The
first invokes gas accretion at the centres 
of galaxies to form stars \citep[e.g.][]{mclaughlin06,nayakshin09}.  
The second postulates successive mergers of globular clusters (GCs) that spiral into
the galactic centre due to dynamical friction \citep[e.g.][]{tremaine75,
quinlan90,gnedin14}.  Several authors have used (primarily N-body) 
simulations to succesfully reproduce some of the observed features of NSCs
formed via this mechanism, including their effective radii and central
velocity dispersions \citep[e.g.][]{capuzzo-dolcetta08}.  For example, 
\citet{gnedin14} recently modeled the effects of dynamical friction on the orbits 
of GCs within their host galaxies, in conjunction 
with the underlying dynamical evolution of the infalling clusters, including 
mass loss (due to both stellar evolution and tidal stripping) and tidal 
disruption.  The model successfully reproduces the observed features of NSCs 
in both late- and early-type galaxies.

Whatever the dominant formation scenario for NSCs, the evidence suggests that
it is an on-going process.  The nuclei of late-type disk galaxies have been
shown to be populated by stars spanning a wide range of ages, with some members
being as young as a few tens of Myrs \citep[e.g.][]{long02}.  Many authors
have suggested that this is indicative of episodic star formation that
has occurred over an extended period of time.  Proponents of the GC infall
scenario, on the other hand, have argued that the young ages of some
of the stars can be accounted for if star clusters spanning a range of
ages were accreted.  In at least the Milky Way, most young star 
clusters are of relatively low-mass in present-day star forming regions.  
On the other hand, \citet{nguyen14} found several super star clusters at the 
centre of Henize 2-10, a blue compact dwarf galaxy with an on-going starburst 
at its centre.  The authors argue that many of these young massive ($\sim$ 
10$^6$ M$_{\odot}$) clusters have very short dynamical friction time-scales, 
suggesting that this may be a rare snapshot of nuclear star cluster formation 
(via the infall of young massive clusters) around a pre-existing super-massive 
black hole.  Regardless, the evidence suggests 
that GC infall cannot be the whole story in at least some NSCs, and that 
in at least these cases some in situ star formation has occurred.  For example, 
this was illustrated by \citet{hartmann11} for the nearby disc galaxies NGC 4244 and 
M33 using both observations and dynamical models of the NSCs at their centres.  
The authors argue that gas dissipation is required to account for $\gtrsim$ 50\% of 
these NSCs' masses.


In this paper, we calculate and compare the mean-square velocities of NSCs and their host 
galaxies, which we use as proxies for energy per unit mass in these systems.  Traditional 
observables, including the radial surface brightness and second-order velocity moment 
profiles, vary on relatively short time-scales due to episodes of matter infall and/or 
star formation.  Meanwhile, the total mass, energy and angular momentum remain approximately 
conserved.  Total energy and angular momentum are decided 
by the various processes that go into forming NSCs and their host galaxies, whereas traditional 
observables are also affected by any subsequent or secondary evolution that occurs in 
energy- and angular momentum-space post-formation.  Thus, the total energy and angular momentum 
more directly probe the formation of galaxies and their nuclear star clusters, by offering 
access to more fundamental properties of the nuclear cluster-galaxy system than 
traditional observables.  However, we caution that our analysis 
relies on the same observable quantities to calculate mean-square velocities and that we 
make the additional 
assumption of virial equilibrium, which introduces some further uncertainty.  
Regardless, as we will illustrate, the comparison is nonetheless instructive.  In particular, 
there is a priori no 
reason to expect that the formation and subsequent time evolution of NSCs should be coupled 
to their host galaxies in energy- and angular momentum-space.  If NSCs and their hosts form 
independently\footnote{For example, while we might naively expect that at least \textit{some} 
correlation should be present in energy-space in most formation scenarios, violent or rapid 
episodes of mass growth in NSCs could erase it.}, then they should remain independent in energy- 
and angular momentum-space 
at the present-day.  Our results suggest that NSCs \textit{are} 
coupled to their hosts in energy-space.  We subsequently explore the physical processes that 
could produce such a correlation and, based on this, propose a means of also using 
total angular momentum to further constrain NSC formation.  
In Section~\ref{background}, we describe the requisite theoretical background.  We present the 
data we use in this paper to calculate the root-mean-square 
velocities in Section~\ref{data}, both for early- and late-type galaxies.  Our results are 
presented in 
Section~\ref{results}.  In Section~\ref{discussion}, we discuss the 
significance of our results for different NSC formation scenarios, and conclude 
in Section~\ref{summary}.

\section{Background} \label{background}

In this section, we describe the necessary theoretical background.  The purpose here 
is to motivate our choice for calculating mean-square velocities for both the central 
NSCs and their host galaxies.  As we will show, the mean-square velocity is a proxy for 
the energy per unit mass of the system, assuming virial equilibrium.  The evolution 
in energy- and angular momentum-space is what determines the balance between the total
system mass, half-mass radius and root-mean-square velocity (see Equations~\ref{eqn:rms1} 
and~\ref{eqn:rms2}), via the virial theorem.  Thus,
the total energy and angular momentum, which are initially determined by the various formation 
processes and are subsequently conserved in time, are fundamental to determining the observed
parameters of a self-gravitating system, which are not time-independent.  Naively, there is no 
reason to expect that the formation and subsequent time evolution of NSCs in energy- and angular 
momentum-space should be coupled to their host galaxies.  If NSCs and their hosts form 
independently, then the evolution in energy- and angular momentum-space should proceed 
independently, first on a crossing time-scale (virialization) and then on a relaxation 
time-scale (two-body relaxation in energy-space and resonant relaxation in angular momentum-space; 
although the two-body relaxation time-scale tends to exceed a Hubble time in 
both NSCs and, especially, their host galaxies) \citep[e.g.][]{merritt13}.  Intriguingly, 
however, we will show that there is a strong (nearly) linear correlation, and in fact a 
near equality, between the energy per unit mass (via the mean-square velocity) of NSCs and 
that of their host galaxies.


We assume virial equilibrium for all NSCs and host galaxies.  The scalar virial relation yields
2E $=$ -2K $=$ W, where E is the total system energy, K is the total kinetic energy and W is the
total potential energy.  The kinetic and potential energies are, respectively:
\begin{equation}
\label{eqn:kin}
K = \frac{1}{2}Mv_{\rm rms}^2
\end{equation}
and
\begin{equation}
\label{eqn:pot}
W = -\frac{GM^2}{r_{\rm g}},
\end{equation}
where v$_{\rm rms}$ and r$_{\rm g}$ are the root-mean-square stellar speed and the gravitational
radius, respectively, and M is the total stellar mass.  It follows from the virial theorem that 
the mean-square velocity is given by:
\begin{equation}
\label{eqn:rms0}
v_{\rm rms}^2 = \frac{|W|}{M} \approx \frac{0.4GM}{r_{\rm h}},
\end{equation}
where the last (approximate) equality holds for many simple stellar systems \citep{spitzer69}, and r$_{\rm h}$ 
is the median radius (or half-mass radius) which is related to the gravitational radius via 
r$_{\rm h}$ $\sim$ 0.4r$_{\rm g}$.  In general, the gravitational radius is related to the system 
half-mass radius via a multiplicative constant that depends on the gravitational potential, or 
r$_{\rm g} =$ $\alpha_{\rm r}$r$_{\rm h}$.  For example, for
a spherically symmetric stellar system, we have r$_{\rm g} =$ 2r$_{\rm h}$ \citep{jaffe83}.
Throughout this paper, we assume r$_{\rm g} =$ 2r$_{\rm h}$ (i.e. $\alpha_{\rm r} =$ 2 in
Equation~\ref{eqn:rms2}) for all NSCs as well as for the bodies of all early-type galaxies, and
r$_{\rm g} =$ r$_{\rm h}$ (i.e. $\alpha_{\rm r} =$ 1) for the bodies of all
late-type galaxies \citep[e.g.][]{binney87} (see below).

We calculate the mean-square velocities for NSCs and their host galaxies using Equations~\ref{eqn:kin}
and Equation~\ref{eqn:pot}.  For the NSCs, we use \citep{binney87}:
\begin{equation}
\label{eqn:rms1}
v_{\rm rms,NSC}^2 = \frac{GM_{\rm NSC}}{2r_{\rm NSC}},
\end{equation}
and for the host galaxies we use:
\begin{equation}
\label{eqn:rms2}
v_{\rm rms,Gal}^2 = \frac{GM_{\rm Gal}}{\alpha_{\rm r}f_{\rm g}r_{\rm Gal}},
\end{equation}
where M$_{\rm NSC}$ and M$_{\rm Gal}$ are, respectively, the total stellar (or baryonic) mass of the NSC 
and host galaxy, f$_{\rm g}$ is the baryonic mass fraction inside r$_{\rm Gal}$, and r$_{\rm NSC}$ 
and r$_{\rm Gal}$ are the effective or projected half-light radii r$_{\rm h}$ of the
NSC and host galaxy, respectively, which we use as proxies for the 3-D half-mass radii (see below).
These masses and radii are derived from the observations as described in Sections~\ref{early} and~\ref{late}.
We further assume a constant dark matter mass fraction (but see Section~\ref{results} for further discussion 
of this issue and its effects on our results) inside the effective radius r$_{\rm h}$ for all
galaxies, corresponding to a baryonic mass fraction f$_{\rm g} = 0.87$ 
\citep{cappellari13a,cappellari13b,toloba14a,toloba14b}.  

In practice, kinematical data (i.e. the two terms in Equation~\ref{eqn:alphav} below, which correspond to
the second-order velocity moments projected along the line of sight) are difficult to obtain.
Consequently, very few such measurements have been provided in the literature
\citep[e.g.][]{busarello92,chae14,toloba14a,toloba14b}.  This is the reason we calculate
root-mean-square velocities indirectly from the virial theorem using the total system mass and
the half-light or effective radius, as opposed to using the directly measured second-order
velocity moments.

How does the root-mean-square velocity calculated from the virial theorem relate to what is 
actually observed?  Similarly to r$_{\rm g}$, the root-mean-square velocity v$_{\rm rms}$ is not 
measured directly, but rather indirectly via the observed second-order velocity moment 
v$^2 +$ $\sigma^2$, where v and $\sigma$ are the observed mean stellar velocity (i.e. the mean velocity 
of ordered bulk motion) and velocity dispersion, respectively.  To 
obtain v$_{\rm rms}^2$ from the observed velocity moments projected along the ling-of-sight, we
would need to include multiplicative factors, or \citep{busarello92}:
\begin{equation}
\label{eqn:alphav}
v_{\rm rms}^2 = \alpha_{\rm v}v^2 + \alpha_{\rm \sigma}\sigma^2
\end{equation}
The constants $\alpha_{\rm v}$ and $\alpha_{\rm \sigma}$ are determined by the gravitational
potential and the angle of inclination relative to the line-of-sight.  For example,
$\alpha_{\rm v} =$ 1 and $\alpha_{\rm \sigma} =$ 3 in the case of an isothermal sphere,
which is spherically symmetric \citep{emsellem07}.  For a non-spherical potential, however,
the angle of inclination must be accounted for when measuring the parameters $\alpha_{\rm v}$
and $\alpha_{\rm \sigma}$.


\section{The data} \label{data}

In this section, we describe the data used to compile our samples of both early- and late-type 
galaxies, beginning with the former.  We summarize in Appendix~\ref{appendix} all NSC 
galaxy masses and root-mean-square velocities calculated from these data.  This is done in 
Tables~\ref{table:table2}, ~\ref{table:table3} and ~\ref{table:table4} 
for each of the late-type, Virgo Cluster and Coma Cluster samples, respectively.

\subsection{Early-type galaxies} \label{early}

For our analysis of early-type galaxies, we use data from the Virgo and Coma Cluster Surveys.  
Beginning with the former, we use data for 47 nucleated early-type galaxies
observed during the
{\it Advanced Camera for Surveys} Virgo Cluster Survey \citep[ACSVCS,][]{cote04}.
We reject five galaxies from the original sample of \citet{cote06} in 
which the apparent NSCs are significantly offset from the galaxy's photocentre and
therefore, as discussed by \citet{cote06}, may well be globular clusters that only
appear to reside close to the nucleus due to a chance projection \citep{leigh12}.  We 
also discard four galaxies with extended nuclei \citep{graham12,scott13}, which are 
more accurately described as nuclear disks \citep{balcells07}, from our sample.  

In order to obtain estimates for the stellar masses, we use the apparent
z-band magnitudes, (g-z) colors, and half-light
radii for both NSCs \citep[from][]{cote06} and host spheroids \citep[from][]{ferrarese06b}.
In order to convert to absolute magnitudes and physical radii, we 
obtain distances to individual galaxies from \citet{tonry01} and \citet{blakeslee02} 
wherever possible, and for galaxies not in the catalogue of \citet{tonry01} we 
adopt a distance of 16.52 Mpc \citep{cote06}.  
We also need to multiply the respective z-band luminosities by an appropriate
mass-to-light ratio.  We use the empirically calibrated mass-to-light
ratios provided by \cite{bell03}, accounting for the (g-z) color of NSC
and spheroid, respectively.  We note here that, given the morphological types of the sample
galaxies (E, S0, dE, dS0, and dE,N), their stellar spheroids can be expected to
be virialized, and to have little or no current star formation activity.
This justifies use of a single (colour-dependent) mass-to-light ratio for
each galaxy spheroid in order to derive its stellar mass.  Approximate error bars for 
the NSC masses were calculated using the 0.041 mag uncertainty quoted by \citet{cote06}.  

We also use data for 53 nucleated low-mass early-type galaxies observed during the 
Coma {\it Advanced Camera for Surveys} Cluster Survey \citep{carter08,denbrok14}.  We 
include all galaxies in the sample designated as possible members or better.  However, 
we also check our derived scaling relations without including galaxies listed as 
possible members, since here membership is not certain.  Our results turn out to be 
insensitive to the adopted inclusion criterion.  To calculate the NSC and 
host galaxy masses, we use the corresponding F814W magnitudes.  Unlike with the Virgo 
Cluster sample, we do not have colour information to calculate empirically-calibrated 
mass-to-light corrections for the stellar population.  Thus, we assume a mass-to-light 
ratio of 2 for all NSCs and host galaxies to correct for the unseen component 
of the stellar mass distribution.  This is a reasonable assumption for at least the 
Coma Cluster, since with only a few exceptions, the colour distribution is 
very narrow \citep{denbrok14}.

\subsection{Late-type galaxies} \label{late}

For our analysis of late-type galaxies, we compiled a sub-sample of 69 galaxies compiled from the sample of
\citet{georgiev14}, for which there are measurements of their PEtrosian radii in the 10th SDSS data 
release\footnote{http://cas.sdss.org/dr10/en/home.aspx}.  These are nearby ($\lesssim$ 40 Mpc), mostly 
low-inclination spirals.  We calculate the total galaxy luminosity and its effective radius using their SDSS 
Petrosian measurements following the prescription given in \citet{graham05} (their Sections 3.2 and 3.3, 
and equations 5 and 6).  To 
obtain masses for these \textit{nuclear star clusters}, we use the luminosities measured from the 
flux within the 
best-fitting King model of a given concentration index (as provided in \citet{georgiev14}),
multiplied by a suitable mass-to-light correction obtained using the \citet{bruzual03} simple
stellar population models for solar metallicity and a \citet{kroupa01} initial mass function.
We use the available colour information given by the various combinations of the most reliable and
well-calculated filters, namely $F450W$, $F555W$, $F606W$ and $F814W$.  All available colours
are matched to the respective colours of the \citet{bruzual03} models, and then a median
mass-to-light ratio is calculated for all colours (for more details see Georgiev et al. 2015,
in preparation).  For the \textit{galaxy} luminosities and sizes, we use the z-band Petrosian magnitudes and 
Petrosian effective radii given in the tenth 
data release of the SDSS database, which correspond 
approximately to the half-light radii of the galaxies.  The luminosities are converted to masses 
using the g- and z-band Petrosian magnitudes along with the \citet{bell03} empirically-calibrated 
mass-to-light ratio colour-corrections.

\section{Results} \label{results}

\begin{figure}
\begin{center}
\includegraphics[width=\columnwidth]{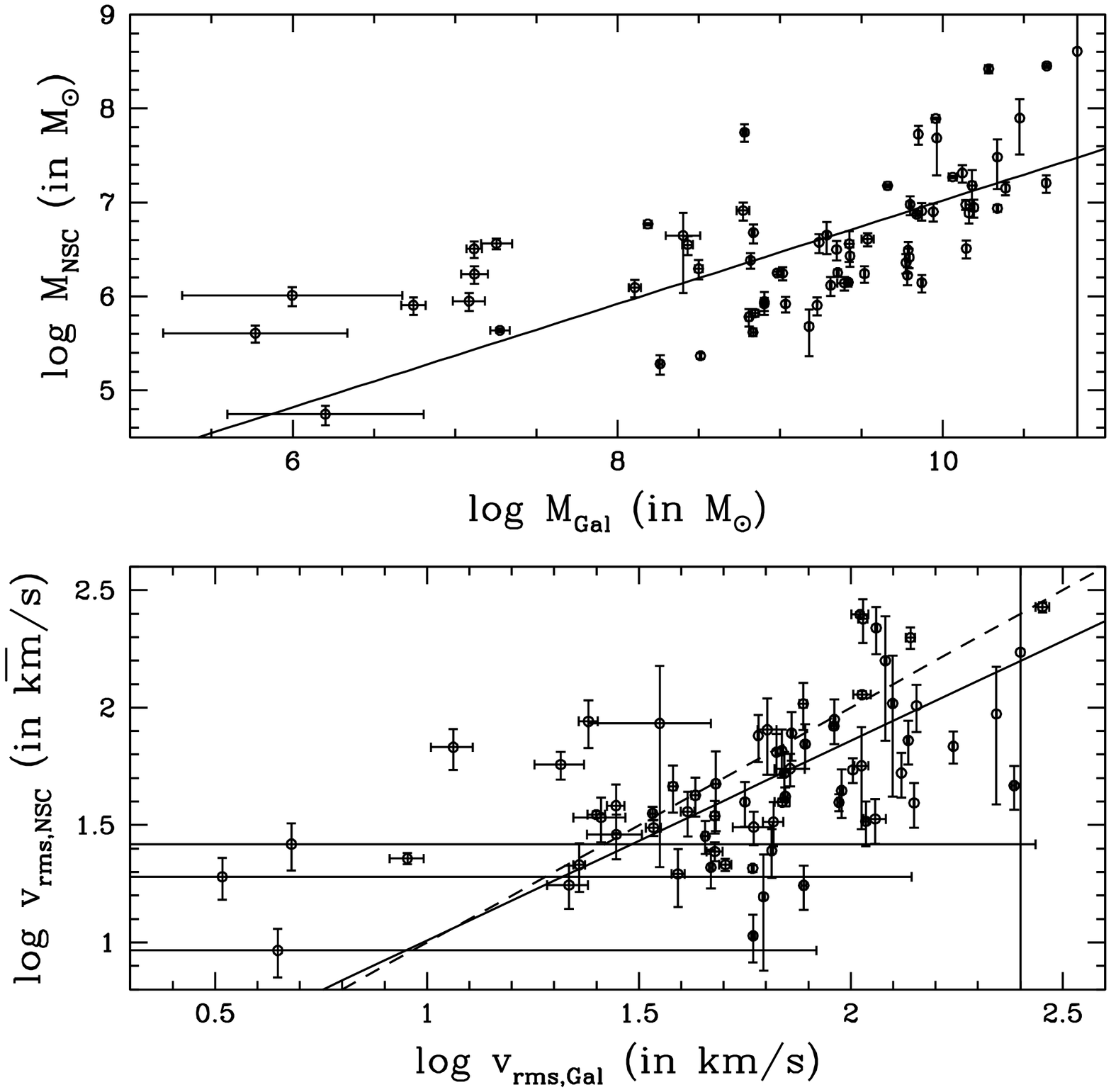}
\end{center}
\caption{The top panel shows the relation between the stellar mass of NSCs and that of 
their host galaxies for 
our sample of late-type galaxies.  The bottom panel shows the relation between the stellar root-mean-square 
velocity for NSCs (Equation~\ref{eqn:rms1}) 
and the root-mean-square velocity (Equations~\ref{eqn:rms2}) of the host galaxy.  
The solid lines show the corresponding lines of best fit given in Table~\ref{table:table1}.  The dashed 
line in the bottom panel shows the one-to-one line.  
\label{fig:fig1}}
\end{figure}

\begin{figure}
\begin{center}
\includegraphics[width=\columnwidth]{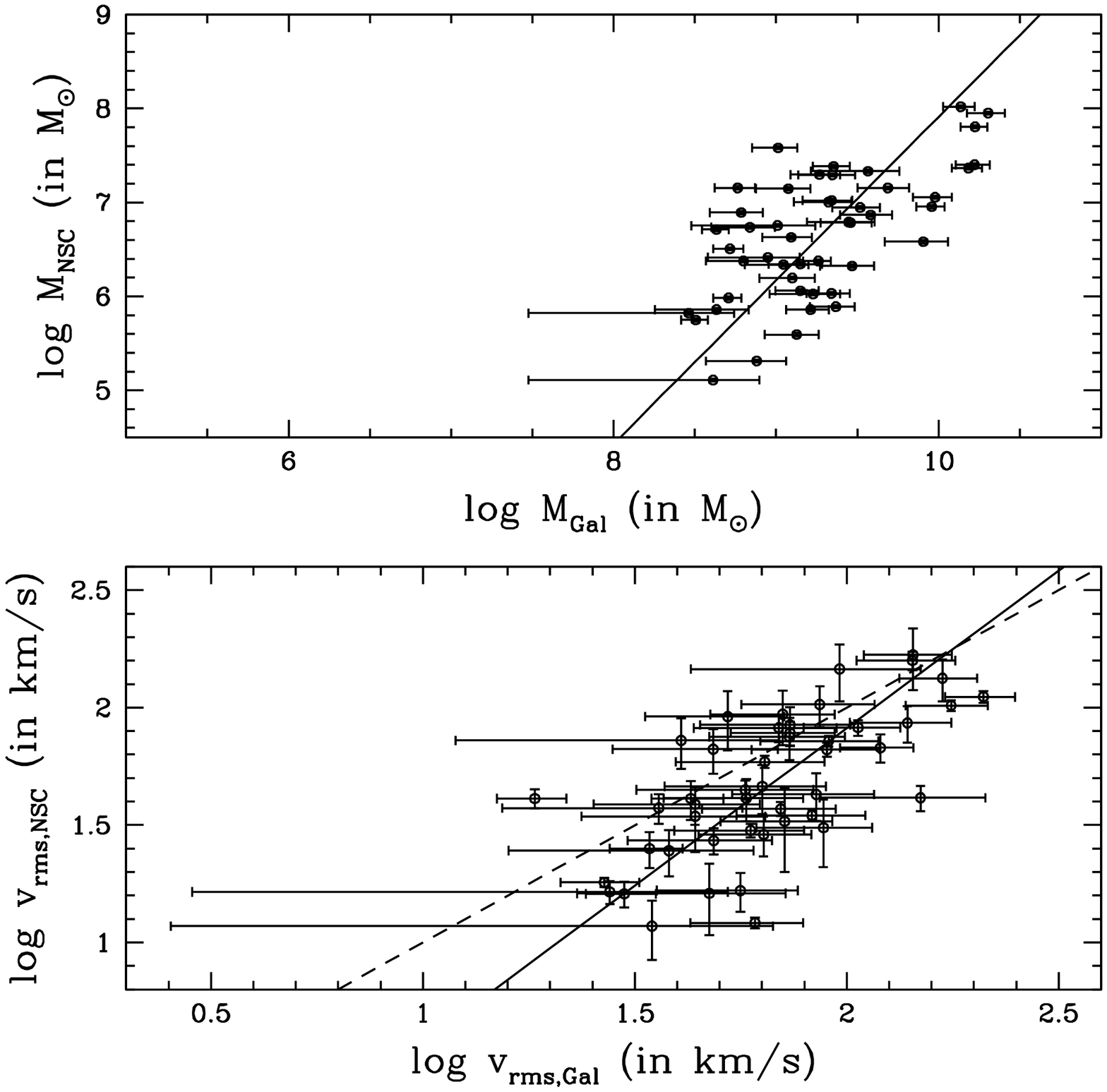}
\end{center}
\caption{Same as Figure~\ref{fig:fig1} but for the Virgo Cluster sample.
\label{fig:fig2}}
\end{figure}

\begin{figure}
\begin{center}
\includegraphics[width=\columnwidth]{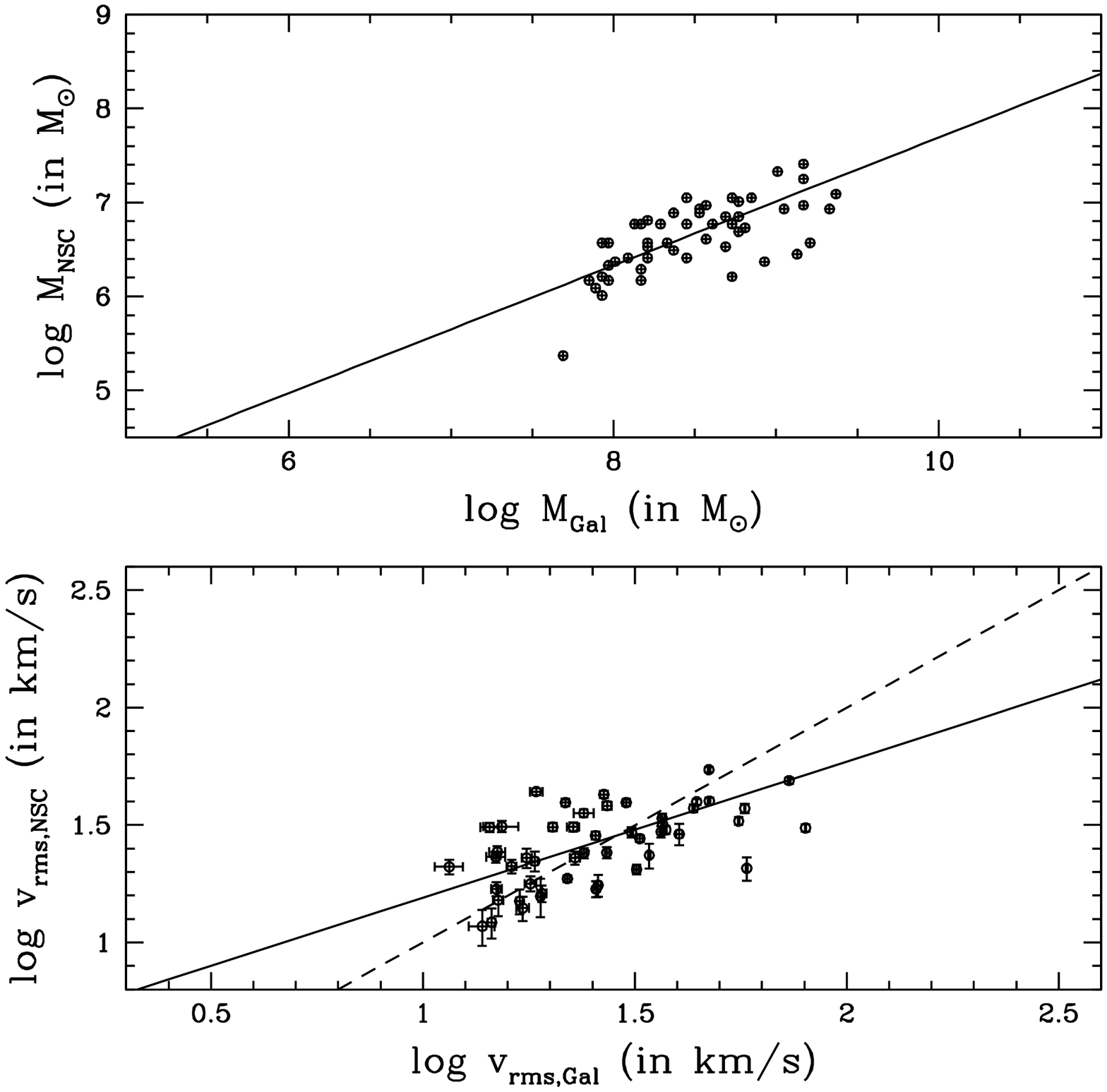}
\end{center}
\caption{Same as Figure~\ref{fig:fig1} but for the Coma Cluster sample.
\label{fig:fig3}}
\end{figure}

The top panels in Figures~\ref{fig:fig1}, ~\ref{fig:fig2} and ~\ref{fig:fig3} show the relation between 
nuclear cluster mass and the mass of its host galaxy for all early- and late-type galaxies.  
The dashed lines have slopes of unity and are shifted along the y-axis to approximately
coincide with the distribution of data points in each panel.  
Lines of best-fit are shown for all samples by the solid lines, found by performing weighted least-square 
fits to the data.  All fit parameters are summarized in Table~\ref{table:table1}.  In all samples, the intrinsic 
dispersion in the data exceeds the uncertainties on 
the individual data points.  This is partly because, in some galaxies, the calculated uncertainties are too small 
to be representative of the true underlying uncertainties in the measurements (see Tables~\ref{table:table2}, 
~\ref{table:table3} and ~\ref{table:table4} in Appendix~\ref{appendix}).  This is 
due to the fact that we 
do not always have errors for both the masses and radii used in our calculations of the root-mean-square 
velocities.  To account for the artificially small error bars, uncertainties on the fit parameters are found 
by adding (in quadrature) 
an additional constant term $\sigma_{\rm int}$ to the uncertainties on both the NSC and host galaxy masses, 
and forcing the resulting reduced chi-square of the fit to be unity.  Hence, the term $\sigma_{\rm int}$ is 
effectively a free parameter in our fits, and is set by the condition $\chi_{\rm red}^2 =$ 1.  This increases 
the uncertainties on the fit parameters 
to more realistically represent the data.  We also divide all galaxy masses by 
10$^9$ M$_{\odot}$, which is approximately equal to the sample means, in order to minimize the 
uncertainties on the y-intercepts of our lines of best-fit.  
For the Virgo Cluster sample we find a slope of 1.74 $\pm$ 0.28 (Virgo), whereas for the Coma Cluster
and late-type samples we find clearly sub-linear
slopes of 0.68 $\pm$ 0.10 (Coma) and 0.55 $\pm$ 0.08 (late-types).  The y-intercepts are 6.47 $\pm$ 0.07
(late-types), 6.17 $\pm$ 0.13 (Virgo) and 7.01 $\pm$ 0.06 (Coma).  
For comparison, we also perform bootstrapped maximum likelihood non-symmetric error-weighted 
fits (without adding an intrisinc dispersion term to the uncertainties on the individual data points).  In all 
but the late-type sample, the derived fit parameters are consistent with our previous estimates to within 
the uncertainties.\footnote{The maximum likelihood method used here is the same as in Georgiev et al. (2015, in preparation), 
where it is described in more detail.  The fit reported here for the late-type sample is taken from Georgiev et al. (2015, 
in preparation), which uses a much larger sample of 247 late-type galaxies.  This accounts for the smaller uncertainties 
on the fit parameters, compared to the Virgo and Coma Cluster samples.}  The slopes for the 
late-type, Virgo and Coma samples are, respectively, 0.55$^{+0.05}_{-0.08}$, 1.92$^{+0.02}_{-0.53}$ and 
0.79$^{+0.08}_{-0.25}$.  The corresponding instrinsic scatter 
estimates are, respectively, 0.211$^{+0.004}_{-0.005}$, 0.203$^{+0.005}_{-0.005}$ and 0.193$^{+0.004}_{-0.005}$.  The 
intrinsic scatter and its uncertainty are found following \citet{jeffreys46} and based on Bayesian probability 
theory, such that the intrinsic dispersion is invariant to rescalings of the problem.

\begin{table*}
\caption{Parameters for all lines of best-fit for the late-type, Virgo and Coma samples.  Each table entry is given 
in the form ($\alpha$; $\beta$), where $\alpha$ and $\beta$ are the slope and y-intercept, respectively.}
\begin{tabular}{|c|c|c|}
\hline
Parameter    &   log M$_{\rm NSC}$ = $\alpha$log (M$_{\rm gal}$/10$^9$) + $\beta$   &   log v$_{rms,NSC}$ = $\alpha$log (v$_{\rm rms,gal}$/40) + $\beta$     \\ 
\hline
Late-type    &   0.55 $\pm$ 0.08; 6.47 $\pm$ 0.07     &  0.85 $\pm$ 0.13; 0.16 $\pm$ 0.25   \\
Virgo        &   1.74 $\pm$ 0.28; 6.17 $\pm$ 0.13     &  1.34 $\pm$ 0.17; -0.78 $\pm$ 0.33  \\
Coma         &   0.68 $\pm$ 0.10; 7.01 $\pm$ 0.06     &  0.58 $\pm$ 0.11; 0.61 $\pm$ 0.16   \\
\hline
\end{tabular}
\label{table:table1}
\end{table*}

As seen in Figure~\ref{fig:fig1}, the fit is poor for the late-type sample.  
This is due to the presence of $\sim$ 10 significant outliers at the low galaxy 
mass end of the distribution, many of which have NSCs that are nearly as massive as their host galaxies.  
Hence, these outliers are in part an artifact of having only included the 
stellar mass in our estimates for the total galaxy masses, and not the gas mass.  In 
late-type galaxies, this effect can be significant.  If the gas mass in these outliers is 
included in the estimates for their total masses, these data points shift to the right in 
Figure~\ref{fig:fig1}, and it becomes apparent that even these NSCs are only $\lesssim$ 1\% the 
mass of their host galaxy.  
Additionally, previous studies have shown that, in at least 
early-type galaxies, the dark matter mass fraction decreases with increasing galaxy mass 
or Sersic index \citep[e.g.][]{cappellari06,forbes08}.  For example, \citet{toloba14b} recently
illustrated that an anti-correlation exists between f$_{\rm DM}$ (within the effective radius) and 
total galaxy luminosity 
for a sample of 39 dwarf early-type galaxies in the Virgo Cluster.  This effect is not included 
in our fits, and should also contribute to increasing the slopes of our lines of best-fit (i.e. closer 
to unity in the Coma Cluster and late-type samples).

The bottom panels in Figures~\ref{fig:fig1}, ~\ref{fig:fig2} and ~\ref{fig:fig3} show the relation between 
nuclear cluster root-mean-square velocity and 
the root-mean-square velocity of its host galaxy, calculated using 
Equations~\ref{eqn:rms1} and~\ref{eqn:rms2}, respectively.  
As before, lines of best-fit (solid lines) are obtained from a weighted least-squares fit to the 
data.  The uncertainties 
on the fit parameters are once again found by accounting for the intrinsic dispersion and 
forcing a reduced chi-square of unity.  In the Virgo Cluster and late-type samples, the slopes are 
consistent with 
being linear to within two standard deviations, or 1.34 $\pm$ 0.17 (Virgo) and 0.85 $\pm$ 0.13 
(late-type), respectively.  In the Coma Cluster sample, however, the slope is sub-linear, or
0.58 $\pm$ 0.11.  With these slopes, we find y-intercepts
of -0.78 $\pm$ 0.33 (Virgo), 0.61 $\pm$ 0.16 (Coma) and 0.16 $\pm$ 0.25 (late-types).
Intriguingly, in both the late-type and Virgo Cluster samples, the slopes shift closer to 
unity relative to the corresponding slopes in Figures~\ref{fig:fig1} and ~\ref{fig:fig2}.
Again, for comparison, we perform bootstrapped maximum likelihood non-symmetric error-weighted 
fits, and find that all of these fits are consistent with those presented above to within one standard 
deviation.  For the late-type, 
Virgo and Coma Cluster samples, we find slopes of, respectively, 
0.96$^{+0.01}_{-0.45}$, 2.13$^{+0.20}_{-1.09}$ and 1.05$^{+0.47}_{-0.99}$.  The corresponding 
instrinsic scatter estimates are, respectively, 0.178$^{+0.002}_{-0.005}$, 0.182$^{+0.004}_{-0.005}$ and 
0.182$^{+0.006}_{-0.004}$.  Note that these intrinsic 
scatter values are comparable to, albeit slightly smaller than, those for the corresponding fits for the 
M$_{\rm gal}$-M$_{\rm NSC}$ relations discussed above.


The bottom panels in Figures~\ref{fig:fig1}, ~\ref{fig:fig2} and ~\ref{fig:fig3} suggest that, to within 
an order of magnitude, NSCs have 
roughly the same root-mean-square velocity, and hence energy per unit mass, as their host galaxies.  
This is the case despite sub-linear slopes for the Coma Cluster sample.  For example, the fit 
for the Coma Cluster sample predicts NSCs at low galaxy masses that have higher root-mean-square 
velocities than 
their hosts.  However, the distribution of galaxy masses, and root-mean-square velocities, is 
narrow for the Coma Cluster sample, so we do not actually observe any NSCs with v$_{\rm rms}$ 
values much higher than their hosts (i.e. by more than order of magnitude).  


\section{Discussion} \label{discussion}

In this section, we discuss the significance of our results for NSC formation.  

\subsection{Energy per unit mass} \label{EpM}

First, we re-iterate a connection between the root-mean-square velocity 
at the half-light radius and the energy per unit mass of the system.  From Equations~\ref{eqn:kin} 
and~\ref{eqn:pot}, 
energy per unit mass scales as E/M $\propto$ v$_{\rm rms}^2 \propto$ M/$\alpha_{\rm r}$r$_{\rm g}$.  
Therefore, if a given NSC has about the same v$_{\rm rms}$ as its host galaxy (within r$_{\rm h}$), 
then both systems have approximately the same energy per unit mass.  As mentioned earlier, this result 
is highly unlikely if NSCs form independently of their host galaxies.  
Our results are consistent with this general picture, since the fit parameters and 
instrinsic scatter we obtain from our statistical analysis suggest that, in all of the late- 
and early-type samples considered here, the relation between NSC root-mean-square velocity and that 
of its host is as statistically significant as the underlying relation between NSC mass and host 
galaxy mass.  However, given the uncertainties inherent to our analysis, this issue should be 
re-visited when better data become available.

Why do NSCs have roughly (to within an order of magnitude) the same energy per unit mass as their host 
galaxy?  Said another way, our results illustrate that the average star in a nuclear star cluster has 
roughly the same 
kinetic energy (and hence total energy, by the virial theorem) as the average non-cluster star
in the host galaxy.  One interpretation of this result is that NSCs were effectively formed
from stars taken directly from the host galaxy, while conserving energy.  Hence, in this
picture, galaxies form first, and then their NSCs subsequently form from some small 
piece of their host galaxy bodies.  From our perspective, this scenario seems to be the 
simplest physically-motivated picture that is consistent with our results.  However, it is by 
no means the only interpretation.  For example, it is conceivable that NSCs and their host 
galaxies form with \textit{different} root-mean-square velocities, and then somehow evolve over 
time toward a common root-mean-square velocity.  However, we are unaware of any physical 
mechanism that might contribute to such a trend. 

\subsection{Implications for NSC formation and evolution} \label{implications}

Our result can be understood within the framework of both the GC infall and in situ star formation models
for NSC formation.  For example, consider the GC infall model for NSC formation in a host galaxy with
constant circular velocity at all radii (i.e. both inside and outside the effective radius r$_{\rm h}$), an 
idealized assumption we will return to below.  The key point is that an infalling 
GC (on roughly a circular orbit) that has just reached the nucleus due to dynamical friction will have 
approximately the same velocity relative 
to the centre of the galaxy as it did several kiloparsecs out.  Hence, when stars orbiting within the GC have
reached the galaxy centre, their velocities (relative to the galaxy centre of mass) remain 
similar to the circular velocity at or beyond $r_{\rm g}$.  It follows that, 
when the inspiraling GC reaches the galaxy's centre and becomes part of the central nuclear
cluster, the energy coming from the bulk motion of the infalling GC is converted to
internal thermal energy within the nuclear cluster itself.  That is, for a stationary observer at the
centre of the galaxy, the centre of mass of the orbiting \textit{cluster} transitions from having a net
velocity roughly equal to the circular velocity to having zero net velocity, whereas the \textit{stars 
within the cluster} always orbit the galaxy centre of mass at the circular velocity.  Thus,
for galaxies with a constant circular velocity inside the effective radius, this predicts a linear 
relationship between the energy per unit mass in the 
central NSC and that of its host galaxy (or, equivalently, a linear relation between the
root-mean-square velocity of the central NSC and that of its host, at all radii inside r$_{\rm g}$).  

Next, consider the in situ star formation scenario for NSC formation, again in a host galaxy with
constant circular velocity at all radii.  Ignoring radiation losses, we 
would naively expect a similar picture as outlined for the GC infall model.  That is, the gas 
should reach the central NSC with the same velocity as it had several kiloparsecs out, creating a 
disk (assuming a preferred plane of accretion) of star-forming gas with a circular velocity that 
roughly matches that of its host (ignoring any central SMBH).  We caution, 
however, that the picture is potentially more complicated within the framework of the in situ star 
formation model for NSC formation.  This is because, when infalling gas filaments collide with the 
forming nucleus, shocks can develop that will radiate energy away and lower the energy per unit mass 
of the central NSC relative to its host.  Thus, at least in the ideal case of a collisionless star-forming 
gas, we expect a central NSC with roughly the same energy per unit mass as its host in galaxies with a 
constant v$_{\rm c}$ at all radii, just as in the GC infall model.

The assumption of a constant circular velocity as a function of galactocentric radius is, to 
first-order, consistent with the observations for most galaxies in our samples (ignoring the innermost 
regions near the galactic centre).  First, 
consider the early-type galaxies.  Observations have now revealed that early-type galaxies are composed of 
both a spherical bulge and an underlying disk \citep[e.g.][]{emsellem07,graham08,scott14,laurikainen11}.
The ratio of bulge-to-disk luminosity in typical
early-type galaxies is L$_{\rm B}$/L$_{\rm D}$ $\sim$ 1/3 \citep[e.g.][]{graham08,laurikainen11}.
Thus, for our early-type samples, the potentials can be approximated by an isothermal density profile for 
the bulge, combined with a Mestel disk \citep{binney87}.  
Subsequently, integrating over the surface brightness profile yields a mass-dependence 
of the form M(r) $\propto$ r.  The key features of these potentials is that they have circular velocities
given by \citep{binney87}:
\begin{equation}
\label{eqn:vcirc}
v_{\rm c} = \frac{GM(r)}{r},
\end{equation}
where r is the distance from the centre of the galaxy, and M(r) is the mass distribution within r.  
The surface brightness is proportional to 1/r in both potentials, hence their sum
also scales as 1/r.  Subsequently, integrating over the surface brightness profile yields a mass-dependence
of the form M(r) $\propto$ r.  From Equation~\ref{eqn:vcirc}, it follows that the circular velocity 
is independent of r in the combined bulge-disk potential, provided 
the outer extent of the bulge extends beyond r$_{\rm h}$.  Beyond the limiting radius of the bulge,
it contributes a constant to the total potential, and v$_{\rm c}$ declines weakly with increasing r.  
For our sample of late-type galaxies, the potential can, to first-order, be approximated by a simple 
Mestel disk, for which the circular velocity is independent of r \citep{binney87}.  Thus, the 
circular velocity is independent of r in this potential as well.  Therefore, in both early-
and late-type galaxies, the assumption of a constant circular velocity v$_{\rm c}$ within the 
effective radius r$_{\rm h}$ is a reasonable first-order assumption.  

We caution that the potentials of \textit{real} galaxies are considerably more complicated than the 
assumptions made above suggest.  In particular, the ratio M(r)/r drops considerably at 
small galactocentric radii, roughly where the NSC potential begins to dominate over that of its 
host.  Previous studies have used numerical simulations to model the properties of NSCs formed from 
GC infall \citep[e.g.][]{antonini12,antonini13}, focusing on the final few parsecs.  \citet{antonini13} 
found that, after many mergers of GCs with the central NSC, the NSC mass-radius relation steepens from 
R$_{\rm NSC}$ $\sim$ M$_{\rm NSC}^{0.5}$ to R$_{\rm NSC}$ $\sim$ M$_{\rm NSC}$.  Thus, the 
energy per unit mass in the NSC ends up being independent of the number of accreted GCs, and 
reaches a roughly constant value.  Using more detailed models, \citet{gnedin14} found a similarly weak 
dependence of the form R$_{\rm NSC}$ $\sim$ M$_{\rm NSC}^{0.23}$.  Therefore, our results suggest a 
similarly weak dependence of the galaxy half-mass radius on the total galaxy stellar mass, in order to 
reproduce the observed relation between the NSC root-mean-square velocity and that of its host.  
Of equal importance, the circular velocity profiles of real galaxies often are not constant at 
large galactocentric radii, but instead show considerable fluctuations 
\citep[e.g.][]{toloba14a,toloba14b}.  Thus, our assumption of a constant circular velocity at all radii 
is not strictly correct, and is likely the source of much of the scatter seen in Figures~\ref{fig:fig1}, 
~\ref{fig:fig2} and ~\ref{fig:fig3}.

What else might a linear relation in energy per unit mass be telling us about NSC formation and/or 
evolution?  In particular, we ask:  How might a linear relation in energy per unit mass between 
NSCs and their host galaxies, once in place, be affected 
by other physical processes characteristic of galaxy formation, such as collisions and mergers 
occurring in galaxy clusters?  
Within the framework of the above picture for NSC formation, which should approximately produce NSCs 
with the same energy per unit mass as their host, one possible interpretation of this result is that 
any collisional events affecting the energy per unit mass of the host must have similarly 
affected its NSC.  In other words, significant energy gains or losses cannot have occurred for 
either the NSC or its host galaxy post-formation.  This is because additional energy deposited within 
NSCs and/or their host galaxies due to direct collisions could 
contribute to increasing the half-light radius of the system, and hence to decreasing the
energy per unit mass (provided the particles involved in the collision are themselves collisionless, which 
is a decent assumption for stars).  This is because the kinetic energies of the colliding galaxies are 
positive and, via energy conservation, must be added to the total energy per unit mass of the 
collision/merger product.  This contributes to a reduction in the total energy per unit mass of the 
system (which is a negative quantity for bound systems), which might be observed as an increase in the 
half-light or effective radius.  Thus, if NSC formation involved more energetic collisions than occurred 
in the host galaxy, this could contribute to a lower root-mean-square
velocity in the NSC relative to its host galaxy.  Conversely, if galaxy formation involved more energetic
collisions than the formation of the central NSC, this could contribute to a lower v$_{\rm rms}$
in the host relative to its central NSC.  If the efficiency of either of these scenarios scales
with total galaxy mass, then this could contribute to a sub-linear or even super-linear dependence
of NSC root-mean-square velocity on host galaxy v$_{\rm rms}$.  In general, collisions that affect 
the NSC more (or less) than its host contribute to a higher (or lower) energy per unit mass in the 
host relative to the central NSC, and should increase the scatter in the bottom panels of 
Figures~\ref{fig:fig1}, ~\ref{fig:fig2} and ~\ref{fig:fig3}.  The intrinsic scatter in 
the observed relations is at least consistent with being due in part to different merger histories 
for the galaxies in our samples.

We caution that the presence of a central SMBH can also strongly affect the evolution and observed 
structural parameters of NSCs.  For example, the presence of a central NSC might be a required
pre-cursor to the formation of an SMBH via core collapse \citep{miller12,gnedin14}.  Subsequently, the
central SMBH can act to disrupt infalling clusters before they reach the nucleus
\citep{antonini12,antonini13}.  Alternatively, a central SMBH or SMBH-SMBH binary could contribute to 
a higher energy per unit mass in NSCs relative to their hosts, by ejecting stars with a large (positive) 
energy per unit mass.  More work is needed to properly quantify the impact of the presence a central 
SMBH on the results presented in this paper.  
Finally, we have also not considered any internal evolution that might occur in NSCs, affecting their 
structural parameters, nor have we considered energy exchange between an existing NSC and its 
host galaxy \citep[e.g.][]{merritt13}.  In general, these omissions are reasonable 
since the relevant time-scales (e.g. two-body relaxation) tend to exceed a Hubble time in
most NSCs \citep[e.g.][]{merritt13}.

\subsection{Future work} \label{future}

We have argued 1) that NSCs form after their host galaxy bodies, and 2) that, in principle, both 
formation scenarios for NSC formation discussed above can produce NSCs with the same energy per unit 
mass as their host galaxies.  In an attempt to identify the most plausible mechanism for NSC 
formation, we now turn our attention to angular momentum.  The general predictions discussed below 
that arise from consideration of angular momentum could be testable in the future with higher 
resolution surveys.

What do we expect for the 
angular momentum content of NSCs relative to their host galaxies?  We 
will argue that, unlike energy per unit mass, the answer could depend on the NSC formation mechanism.  
In particular, GC orbits form a halo within the host galaxy, with their orbital planes aligned at random 
angles relative to each other.  Thus, if only 
a single GC inspiral is responsible for forming the NSC, then we would expect significant rotation 
in the remnant NSC, with the axis of rotation aligned perpendicular to the original orbital plane 
of the inspiraling GC.  This rotation can be canceled by the cumulative effects of many GCs inspiralling into 
the nucleus, which should create a central NSC with roughly the same energy per unit mass as its host, but 
nearly zero net angular momentum (within the framework of the picture described above).\footnote{In 
fact, repeated episodes of GC infall may still yield non-negligible rotation in the NSC, provided the orbits 
characteristic of the bulk of the GC population in the host galaxy exhibit a clear rotation signature, 
with v/$\sigma$ $>$ 0.5.  This has been observed in a number of early-type galaxies 
\citep[e.g.][]{beasley09,pota13}, and even the bulge of the Milky Way \citep[e.g.][]{harris96}.}  
This is the case for both early- and late-type galaxies.  The story changes within the framework 
of the in situ star formation model for NSC growth.  This is because the 
cumulative effects of galaxy mergers have not depleted the total angular momentum content of 
late-type galaxies.  In late-types, there \textit{is} a preferred plane of accretion for gas, but 
not for any accreted GCs.  Therefore, any in situ star formation occurring in a NSC forming via 
gas accretion should generally occur in a 
disk, with its dominant angular momentum vector aligned with that of the host.  NSCs formed 
from many episodes of GC infall, however, should have roughly zero net angular momentum.  Thus, we 
predict that 
NSCs in late-type galaxies formed primarily via gas accretion and in situ star formation
should contain significant angular momentum, which is directly observable as rotation.  Naively, this 
also predicts a correlation between the 
observed inclination of the host galaxy disk relative to the observer line of sight and
the (isophotal) ellipticity of the NSC, with more edge-on disk galaxies having more flattened NSCs.  
In early-type galaxies, on the other hand, there is no preferred plane of accretion for the gas.  
Thus, we expect a central NSC with roughly the same energy per unit mass as its host, just as in the GC
infall model, but with little to no net angular momentum or rotation.  

The implication that NSCs are formed and evolve together with their host galaxies suggests that the 
origins of NSCs can be distinguished observationally (again, ignoring any internal evolution occuring 
within the NSCs) 
when better resolution becomes available.  This can be done by decomposing NSCs into bulge and disk
components, and comparing the bulge and disk masses, as
well as the angle of inclination between the plane of the NSC disk and that of its host
galaxy.  This is especially true in spiral galaxies, since they
have a milder merger history than ellipticals.  Consider the quantity
$\beta =$ cos${\rm \theta}$M$_{\rm disk}$/(M$_{\rm bulge}+$M$_{\rm disk}$), where M$_{\rm disk}$ is the
mass of the disk component,
M$_{\rm bulge}$ is the mass of the bugle component and ${\rm \theta}$ is the angle of inclination
between the NSC disk and that of its host galaxy.  If $\beta \lesssim$ 0, then the NSC was formed mainly
through GC infall.  If, on the other hand, $\beta > 0$ then a non-negligible fraction of the NSC
formed through in situ star formation, with $\beta =$ 1 corresponding to the case of 0\% GC infall
and 100\% in situ star formation.


\section{Summary} \label{summary}

We estimate and compare the root-mean-square velocities for a large sample of 
nuclear star clusters and their host galaxies (both early- and late-type).  These are used 
as proxies for energy per unit mass, and it is 
demonstrated that NSCs have roughly the same energy per unit mass as their host galaxies, to 
within an order of magnitude.  The origin of this interesting relation is discussed.  
We interpret this as evidence that NSCs do not form independently
of their host galaxies, but rather that their formation and subsequent evolution are coupled.
We discuss how our results can potentially be used to offer a clear and observationally testable
prediction to distinguish between the different nuclear star cluster formation scenarios, and even
quantify their relative contributions.

\appendix 

\section{Data} \label{appendix}

In this appendix, we present the data used to generate Figures~\ref{fig:fig1}, ~\ref{fig:fig2} 
and ~\ref{fig:fig3}.  All calculated NSC and galaxy masses and root-mean-square 
velocities are summarized in Tables~\ref{table:table2}, ~\ref{table:table3} and ~\ref{table:table4} 
for each of the late-type, Virgo Cluster and Coma Cluster samples, respectively.

\begin{table*}
\caption{Properties of all late-type galaxies.  Column 1 gives the object ID.  
Columns 2 and 3 list the total stellar mass of the galaxy and NSC in units of 10$^9$ M$_{\odot}$ and 
10$^6$ M$_{\odot}$, respectively.  
Columns 4 and 5 provide the root-mean-square velocities for the galaxy and NSC, respectively, in units 
of km s$^{-1}$.  The galaxy and NSC mass uncertainties are taken directly from the literature.  The uncertainties 
for the root-mean-square velocities are calculated using the provided uncertainties on both the 
galaxy or NSC masses and radii, wherever available.}
\begin{tabular}{|c|c|c|c|c|}
\hline
ID    &   M$_{\rm gal}$  &   M$_{\rm NSC}$    &   v$_{\rm rms,gal}$   &    v$_{\rm rms,NSC}$      \\
      &  (10$^9$ M$_{\odot}$)   &  (10$^6$ M$_{\odot}$)  &    (km s$^{-1}$)  &  (km s$^{-1}$)  \\
\hline
IC0769          &      8.761$^{0.099}_{-0.098}$   &     7.963 $\pm$ 1.71      &  78.12 $\pm$  0.7954    &  69.95 $\pm$  15.02  \\
M108            &      0.127$^{0.011}_{-0.010}$   &     1.239 $\pm$ 0.266     &  65.68 $\pm$   3.675    &  32.65 $\pm$  7.009  \\
MCG-01-03-085   &      0.658$^{0.013}_{-0.013}$   &     2.417 $\pm$ 0.4541    &  42.96 $\pm$  0.8615    &  42.34 $\pm$  7.954  \\
NGC0428         &      2.661$^{0.151}_{-0.143}$   &     3.619 $\pm$ 1.297     &  63.53 $\pm$   3.214    &  80.54 $\pm$  28.87  \\
NGC0600         &      6.232$^{0.051}_{-0.051}$   &     2.595 $\pm$ 0.6061    &  68.75 $\pm$  0.5458    &  65.52 $\pm$   15.3  \\
NGC0853         &      2.691$^{0.021}_{-0.021}$   &     2.692 $\pm$ 0.6289    &  69.64 $\pm$  0.6369    &  52.51 $\pm$  12.26  \\
NGC0864         &       9.07$^{0.451}_{-0.430}$   &     78.0  $\pm$ 2.024     &  105.1 $\pm$   4.838    &  249.3 $\pm$  6.466  \\
NGC1042         &      0.253$^{0.070}_{-0.055}$   &     4.432 $\pm$ 3.346     &  35.41 $\pm$   11.38    &  85.62 $\pm$  64.64  \\
NGC2500         &      0.800$^{0.030}_{-0.029}$   &    0.8753 $\pm$ 0.2416    &  39.13 $\pm$   1.374    &   19.6 $\pm$   5.41  \\
NGC2541         &      0.643$^{0.027}_{-0.026}$   &    0.6041 $\pm$ 0.1297    &  41.26 $\pm$   1.527    &  36.04 $\pm$  7.738  \\
NGC2552         &      0.270$^{0.021}_{-0.019}$   &     3.549 $\pm$ 0.8097    &  24.06 $\pm$   1.237    &  87.37 $\pm$  19.93  \\
NGC2805         &      6.314$^{0.120}_{-0.118}$   &     9.509 $\pm$ 2.169     &  77.24 $\pm$   1.515    &  103.7 $\pm$  23.67  \\
NGC3041         &      15.17$^{0.677}_{-0.648}$   &     15.15 $\pm$ 6.995     &  105.9 $\pm$   4.083    &  56.52 $\pm$  26.09  \\
NGC3259         &      13.23$^{0.085}_{-0.085}$   &     20.56 $\pm$ 4.327     &  136.6 $\pm$  0.8513    &  72.56 $\pm$  15.27  \\
NGC3274         &      0.681$^{0.023}_{-0.023}$   &    0.4157 $\pm$ 0.0390    &  47.78 $\pm$   2.064    &  24.41 $\pm$  2.293  \\
NGC3319         &      1.036$^{0.014}_{-0.014}$   &     1.759 $\pm$ 0.2793    &  45.37 $\pm$   0.688    &  28.37 $\pm$  4.505  \\
NGC3338         &      19.23$^{0.455}_{-0.444}$   &     263.9 $\pm$ 27.08     &  138.3 $\pm$   3.337    &  198.5 $\pm$  20.37  \\
NGC3346         &      6.142$^{0.055}_{-0.055}$   &     3.091 $\pm$ 0.722     &  72.56 $\pm$  0.4986    &  77.74 $\pm$  18.16  \\
NGC3359         &      13.88$^{0.043}_{-0.043}$   &     9.465 $\pm$ 1.153     &  101.1 $\pm$  0.2946    &  54.31 $\pm$  6.618  \\
NGC3423         &      2.263$^{0.029}_{-0.029}$   &     1.791 $\pm$ 0.1636    &  70.02 $\pm$  0.9634    &  41.84 $\pm$  3.823  \\
NGC3445         &      1.736$^{0.009}_{-0.009}$   &     3.751 $\pm$ 0.8557    &  60.51 $\pm$  0.2824    &  75.91 $\pm$  17.32  \\
NGC3455         &      7.435$^{0.049}_{-0.049}$   &     8.111 $\pm$ 1.741     &  91.52 $\pm$  0.5246    &  89.05 $\pm$  19.12  \\
NGC3666         &      9.215$^{0.063}_{-0.063}$   &     48.36 $\pm$ 28.98     &  125.6 $\pm$  0.9413    &  104.1 $\pm$  62.37  \\
NGC3756         &      21.71$^{0.119}_{-0.118}$   &     30.35 $\pm$ 16.53     &  120.7 $\pm$  0.5633    &  158.4 $\pm$  86.31  \\
NGC3913         &      2.046$^{0.018}_{-0.018}$   &     1.314 $\pm$ 0.3015    &   58.8 $\pm$  0.6037    &  10.67 $\pm$   2.45  \\
NGC3949         &      14.53$^{0.017}_{-0.017}$   &     7.723 $\pm$ 1.765     &  142.9 $\pm$  0.1635    &  101.9 $\pm$  23.29  \\
NGC4030         &      43.76$^{1.497}_{-1.447}$   &     284.0 $\pm$ 14.45     &  283.4 $\pm$   10.26    &  268.1 $\pm$  13.64  \\
NGC4041         &      29.84$^{0.050}_{-0.050}$   &     78.82 $\pm$ 46.31     &  220.6 $\pm$  0.4428    &  93.96 $\pm$  55.21  \\
NGC4062         &      13.98$^{0.025}_{-0.025}$   &     3.229 $\pm$ 0.6933    &  131.7 $\pm$  0.2014    &  52.71 $\pm$  11.32  \\
NGC4096         &      7.449$^{0.104}_{-0.103}$   &     1.397 $\pm$ 0.2999    &  108.5 $\pm$   1.624    &  32.75 $\pm$  7.032  \\
NGC4204         &      0.182$^{0.006}_{-0.006}$   &    0.1915 $\pm$ 0.0447    &  22.89 $\pm$  0.7458    &  21.39 $\pm$  4.996  \\
NGC4208         &      15.57$^{0.057}_{-0.057}$   &     8.812 $\pm$ 1.892     &   141. $\pm$  0.4547    &  39.25 $\pm$  8.427  \\
NGC4237         &      24.37$^{0.085}_{-0.085}$   &     14.15 $\pm$ 2.209     &  174.5 $\pm$  0.5642    &  68.42 $\pm$  10.68  \\
NGC4393         &      0.001$^{0.002}_{-0.000}$   &    0.4048 $\pm$ 0.0819    &  3.289 $\pm$    136.    &  19.05 $\pm$  3.852  \\
NGC4395         &      0.013$^{0.001}_{-0.001}$   &     3.203 $\pm$ 0.6301    &  11.53 $\pm$   1.301    &  67.77 $\pm$  13.33  \\
NGC4411b        &      0.684$^{0.015}_{-0.015}$   &     4.766 $\pm$ 1.087     &  38.07 $\pm$  0.7635    &  46.21 $\pm$  10.54  \\
NGC4498         &      0.013$^{0.003}_{-0.002}$   &     1.726 $\pm$ 0.3706    &  25.77 $\pm$   3.636    &  34.06 $\pm$  7.313  \\
NGC4517         &      0.012$^{0.003}_{-0.002}$   &    0.8894 $\pm$ 0.1909    &  27.99 $\pm$   4.157    &  28.84 $\pm$  6.191  \\
NGC4522         &       6.07$^{0.063}_{-0.063}$   &     1.679 $\pm$ 0.3605    &  77.42 $\pm$   1.023    &  17.49 $\pm$  3.756  \\
NGC4525         &      3.307$^{0.018}_{-0.018}$   &     1.745 $\pm$ 0.3451    &  66.77 $\pm$  0.3455    &  64.58 $\pm$  12.77  \\
NGC4534         &      1.079$^{0.014}_{-0.014}$   &    0.8319 $\pm$ 0.1557    &  46.78 $\pm$  0.5342    &  20.89 $\pm$   3.91  \\
NGC4567         &      0.604$^{0.018}_{-0.017}$   &      55.6 $\pm$ 11.7      &  106.8 $\pm$   2.657    &  238.6 $\pm$  50.21  \\
NGC4571         &      6.938$^{0.103}_{-0.101}$   &     7.449 $\pm$ 0.1933    &  91.27 $\pm$   1.382    &  83.46 $\pm$  2.165  \\
NGC4592         &       1.69$^{0.007}_{-0.007}$   &    0.8037 $\pm$ 0.1725    &  56.28 $\pm$  0.2334    &  39.64 $\pm$   8.51  \\
NGC4618         &      1.507$^{0.010}_{-0.010}$   &    0.4788 $\pm$ 0.2468    &  62.28 $\pm$  0.3859    &  15.66 $\pm$  8.069  \\
NGC4625         &      0.957$^{0.004}_{-0.004}$   &     1.772 $\pm$ 0.0599    &  58.72 $\pm$  0.2616    &  20.69 $\pm$  0.699  \\
NGC4631         &      0.006$^{0.001}_{-0.001}$   &    0.8037 $\pm$ 0.1691    &  21.62 $\pm$   2.386    &  17.57 $\pm$  3.696  \\
NGC4635         &      1.937$^{0.035}_{-0.035}$   &     4.488 $\pm$ 1.668     &  48.05 $\pm$   0.719    &  47.38 $\pm$  17.61  \\
NGC4651         &      67.51$^{0.199}_{-0.199}$   &     404.8 $\pm$ 93210000. &  251.2 $\pm$  0.721     &  172.1 $\pm$  39620000.  \\
NGC4701         &      7.094$^{0.044}_{-0.044}$   &     53.23 $\pm$ 12.14     &  114.9 $\pm$  0.665     &  218.4 $\pm$  49.82  \\
NGC4900         &      11.55$^{0.707}_{-0.666}$   &     18.59 $\pm$ 0.6053    &  106.4 $\pm$   5.072    &  113.6 $\pm$  3.698  \\
NGC4904         &      5.951$^{0.055}_{-0.054}$   &     2.281 $\pm$ 0.5328    &  95.14 $\pm$  0.857     &   44.3 $\pm$  10.35  \\
NGC5112         &      2.633$^{0.097}_{-0.094}$   &      1.41 $\pm$ 0.0859    &  50.58 $\pm$   1.697    &  21.43 $\pm$  1.306  \\
NGC5204         &      0.324$^{0.004}_{-0.004}$   &    0.2335 $\pm$ 0.0153    &  34.09 $\pm$  0.4149    &  35.43 $\pm$   2.32  \\
NGC5300         &      4.591$^{0.203}_{-0.195}$   &     15.03 $\pm$ 0.5514    &   68.9 $\pm$   2.784    &  39.61 $\pm$  1.453  \\
NGC5334         &      21.76$^{0.218}_{-0.216}$   &     8.635 $\pm$ 0.716     &  93.79 $\pm$    0.77    &   39.5 $\pm$  3.276  \\
NGC5585         &      0.799$^{0.017}_{-0.016}$   &    0.8341 $\pm$ 0.1325    &  47.78 $\pm$  0.964     &  34.58 $\pm$  5.492  \\
NGC5668         &      3.452$^{0.308}_{-0.283}$   &     4.066 $\pm$ 0.6458    &  71.93 $\pm$   5.838    &  54.91 $\pm$  8.721  \\
NGC5964         &      0.018$^{0.004}_{-0.004}$   &     3.648 $\pm$ 0.4912    &   20.7 $\pm$   2.778    &  57.18 $\pm$  7.698  \\
NGC5970         &      0.592$^{0.056}_{-0.051}$   &     8.186 $\pm$ 1.758     &  114.2 $\pm$   6.918    &  33.59 $\pm$  7.212  \\
NGC6384         &      43.43$^{0.386}_{-0.383}$   &     16.04 $\pm$ 3.374     &  242.8 $\pm$   2.865    &  46.57 $\pm$  9.799  \\
NGC7741         &      2.487$^{0.260}_{-0.235}$   &     1.384 $\pm$ 0.2264    &    59. $\pm$   6.378    &  30.99 $\pm$  5.068  \\
UGC02302        &      0.019$^{0.003}_{-0.002}$   &    0.4345 $\pm$ 0.0232    &  8.993 $\pm$  0.821     &  22.78 $\pm$  1.218  \\
UGC03860        &      0.002$^{0.005}_{-0.001}$   &    0.0557 $\pm$ 0.0130    &  4.445 $\pm$   78.45    &  9.253 $\pm$  2.161  \\
UGC06192        &      0.315$^{0.018}_{-0.017}$   &     1.969 $\pm$ 0.4598    &  27.92 $\pm$   1.332    &  38.21 $\pm$  8.924  \\
UGC06983        &      0.701$^{0.030}_{-0.028}$   &    0.6627 $\pm$ 0.0507    &  34.25 $\pm$   1.377    &  30.82 $\pm$  2.357  \\
UGC08041        &      0.154$^{0.009}_{-0.008}$   &     5.878 $\pm$ 0.1914    &  25.11 $\pm$   1.204    &  35.03 $\pm$  1.141  \\
UGC08516        &      2.226$^{0.015}_{-0.015}$   &     3.149 $\pm$ 0.7355    &  65.14 $\pm$  0.406     &  24.59 $\pm$  5.743  \\
UGC12732        &      0.001$^{0.004}_{-0.001}$   &     1.024 $\pm$ 0.2335    &  4.787 $\pm$    268.    &  26.23 $\pm$  5.984  \\
\hline
\end{tabular}
\label{table:table2}
\end{table*}

\begin{table*}
\caption{Properties of all Virgo Cluster galaxies.  The columns are the same as in Table~\ref{table:table2}.}
\begin{tabular}{|c|c|c|c|c|}
\hline
ID       &    M$_{\rm gal}$  &   M$_{\rm NSC}$    &   v$_{\rm rms,gal}$   &    v$_{\rm rms,NSC}$      \\
         &   (10$^9$ M$_{\odot}$)   &  (10$^6$ M$_{\odot}$)  &    (km s$^{-1}$)  &  (km s$^{-1}$)  \\
\hline
VCC1720    &  20.21 $\pm$ 5.31   &  89.24 $\pm$ 3.371     &    142.9 $\pm$ 37.55    &    158.5 $\pm$ 14.36 \\
VCC1883    &  16.66 $\pm$ 3.92   &  25.12 $\pm$ 0.949     &    143.4 $\pm$ 33.73    &    167.7 $\pm$ 49.31 \\
VCC1242    &  15.27 $\pm$ 3.17   &  23.14 $\pm$ 0.8741    &    168.1 $\pm$ 34.91    &    133.2 $\pm$ 27.12 \\
VCC784     &  16.79 $\pm$ 3.14   &  63.61 $\pm$ 2.403     &    210.1 $\pm$ 39.29    &    111.1 $\pm$ 6.401 \\
VCC828     &  13.69 $\pm$ 3.00   &  104.4 $\pm$ 3.942     &    176.3 $\pm$ 38.64    &    102.  $\pm$ 5.162 \\
VCC1250    &   3.68 $\pm$ 2.04   &  21.48 $\pm$ 0.8115    &    96.18 $\pm$ 53.32    &    145.9 $\pm$ 39.67 \\
VCC1125    &   8.06 $\pm$ 3.4    &  3.814 $\pm$ 0.144     &    149.3 $\pm$ 62.98    &    41.31 $\pm$ 5.066 \\
VCC1283    &   9.08 $\pm$ 1.79   &   8.99 $\pm$ 0.3396    &    120.  $\pm$ 23.66    &    67.49 $\pm$ 9.271 \\
VCC1261    &   4.87 $\pm$ 1.69   &  14.25 $\pm$ 0.5383    &    86.41 $\pm$ 29.99    &    103.1 $\pm$ 20.42 \\
VCC698     &   9.52 $\pm$ 2.56   &  11.33 $\pm$ 0.4278    &    139.  $\pm$ 37.39    &    86.12 $\pm$ 15.06 \\
VCC1422    &   3.82 $\pm$ 1.35   &  7.342 $\pm$ 0.2773    &    73.24 $\pm$ 25.88    &    75.05 $\pm$ 15.28 \\
VCC2048    &   2.94 $\pm$ 1.08   &  2.105 $\pm$ 0.0795    &    84.73 $\pm$ 31.12    &    42.7  $\pm$ 9.777 \\
VCC1871    &   2.26 $\pm$ 0.58   &  24.18 $\pm$ 0.9134    &    106.3 $\pm$ 27.29    &    82.14 $\pm$ 6.162 \\
VCC1910    &    2.1 $\pm$ 0.81   &   10.1 $\pm$ 0.3814    &    73.46 $\pm$ 28.34    &    84.47 $\pm$ 15.88 \\
VCC856     &   2.22 $\pm$ 0.85   &  19.66 $\pm$ 0.7427    &    64.05 $\pm$ 24.53    &    58.74 $\pm$ 3.485 \\
VCC140     &   2.33 $\pm$ 0.71   & 0.7763 $\pm$ 0.0293    &    88.08 $\pm$ 26.84    &    30.78 $\pm$ 9.863 \\
VCC1355    &   1.82 $\pm$ 0.35   &   2.38 $\pm$ 0.0899    &    42.9  $\pm$ 8.25     &    41.01 $\pm$ 7.712 \\
VCC1087    &   3.29 $\pm$ 1.07   &  8.816 $\pm$ 0.333     &    70.6  $\pm$ 22.96    &    93.64 $\pm$ 24.53 \\
VCC1861    &   2.88 $\pm$ 1.00   &   6.08 $\pm$ 0.2296    &    69.81 $\pm$ 24.24    &    37.04 $\pm$ 2.589 \\
VCC543     &   2.19 $\pm$ 0.65   &  1.074 $\pm$ 0.0406    &    60.79 $\pm$ 18.04    &    12.13 $\pm$ 0.631 \\
VCC1431    &    2.2 $\pm$ 0.74   &  10.49 $\pm$ 0.3963    &    82.82 $\pm$ 27.86    &    34.77 $\pm$ 1.678 \\
VCC1528    &   1.63 $\pm$ 0.48   & 0.7213 $\pm$ 0.0272    &    71.36 $\pm$ 21.01    &    32.8  $\pm$ 12.82 \\
VCC1695    &   1.69 $\pm$ 0.78   &   1.06 $\pm$ 0.0400    &    43.91 $\pm$ 20.27    &    34.44 $\pm$ 10.13 \\
VCC437     &    2.8 $\pm$ 1.25   &   6.18 $\pm$ 0.2334    &    57.66 $\pm$ 25.74    &    44.71 $\pm$ 4.132 \\
VCC2019    &   1.02 $\pm$ 0.72   &  5.671 $\pm$ 0.2142    &    40.64 $\pm$ 28.69    &    72.46 $\pm$ 17.7  \\
VCC33      &   0.43 $\pm$ 0.25   & 0.7211 $\pm$ 0.0272    &    38.06 $\pm$ 22.13    &    24.6  $\pm$ 5.461 \\
VCC200     &   1.34 $\pm$ 0.49   &  0.392 $\pm$ 0.0148    &    56.15 $\pm$ 20.53    &    16.64 $\pm$ 3.13  \\
VCC1488    &   0.41 $\pm$ 0.38   & 0.1285 $\pm$ 0.0049    &    34.72 $\pm$ 32.18    &    11.75 $\pm$ 3.319 \\
VCC1895    &   0.76 $\pm$ 0.39   & 0.2053 $\pm$ 0.0078    &    47.42 $\pm$ 24.34    &    16.2  $\pm$ 5.435 \\
VCC1545    &   1.41 $\pm$ 0.42   &  1.146 $\pm$ 0.0433    &    63.66 $\pm$ 18.96    &    28.84 $\pm$ 5.563 \\
VCC1192    &   1.85 $\pm$ 0.62   &  19.78 $\pm$ 0.747     &    89.74 $\pm$ 30.08    &    66.25 $\pm$ 4.577 \\
VCC1075    &   1.11 $\pm$ 0.47   &  2.178 $\pm$ 0.0823    &    43.87 $\pm$ 18.58    &    38.73 $\pm$ 7.103 \\
VCC1627    &   1.03 $\pm$ 0.32   &  38.08 $\pm$ 1.438     &    90.63 $\pm$ 28.16    &    72.04 $\pm$ 3.736 \\
VCC1440    &   1.19 $\pm$ 0.44   &    14. $\pm$ 0.5287    &    69.07 $\pm$ 25.54    &    81.93 $\pm$ 10.7  \\
VCC230     &   0.69 $\pm$ 0.29   &  5.445 $\pm$ 0.2057    &    48.38 $\pm$ 20.33    &    66.56 $\pm$ 14.34 \\
VCC2050    &   0.29 $\pm$ 0.26   & 0.6627 $\pm$ 0.0250    &    27.61 $\pm$ 24.76    &    16.42 $\pm$ 1.849 \\
VCC751     &   1.41 $\pm$ 0.51   &  2.203 $\pm$ 0.0832    &    57.9  $\pm$ 20.94    &    41.12 $\pm$ 8.369 \\
VCC1828    &   1.26 $\pm$ 0.47   &  1.567 $\pm$ 0.0592    &    48.56 $\pm$ 18.11    &    27.17 $\pm$ 3.491 \\
VCC538     &   0.63 $\pm$ 0.26   &  2.382 $\pm$ 0.0900    &    63.25 $\pm$ 26.1     &    46.17 $\pm$ 10.91 \\
VCC1407    &   1.24 $\pm$ 0.42   &  4.252 $\pm$ 0.1606    &    59.28 $\pm$ 20.08    &    29.98 $\pm$ 2.003 \\
VCC1886    &   0.51 $\pm$ 0.1    & 0.9599 $\pm$ 0.0363    &    34.27 $\pm$ 6.72     &    25.07 $\pm$ 4.384 \\
VCC1199    &   0.58 $\pm$ 0.16   &  14.29 $\pm$ 0.5396    &    73.43 $\pm$ 20.26    &    78.03 $\pm$ 9.158 \\
VCC1539    &   0.52 $\pm$ 0.11   &  3.209 $\pm$ 0.1212    &    26.78 $\pm$ 5.665    &    18.03 $\pm$ 0.831 \\
VCC1185    &   0.89 $\pm$ 0.51   &    2.6 $\pm$ 0.0982    &    36.03 $\pm$ 20.65    &    37.36 $\pm$ 5.418 \\
VCC1826    &   0.61 $\pm$ 0.22   &  7.802 $\pm$ 0.2947    &    52.35 $\pm$ 18.88    &    91.54 $\pm$ 25.86 \\
VCC1489    &   0.32 $\pm$ 0.06   & 0.5612 $\pm$ 0.0212    &    29.84 $\pm$ 5.596    &    16.12 $\pm$ 2.038 \\
VCC1661    &   0.43 $\pm$ 0.08   &  5.142 $\pm$ 0.1942    &    18.38 $\pm$ 3.42     &    41.03 $\pm$ 3.83  \\
\hline
\end{tabular}
\label{table:table3}
\end{table*}

\begin{table*}
\caption{Properties of all Coma Cluster galaxies.  The columns are the same as in Table~\ref{table:table2}.  Uncertainties 
on the NSC masses are provided when available.}
\begin{tabular}{|c|c|c|c|c|}
\hline
ID        &    M$_{\rm gal}$  &   M$_{\rm NSC}$    &   v$_{\rm rms,gal}$   &    v$_{\rm rms,NSC}$      \\
          &   (10$^9$ M$_{\odot}$)   &  (10$^6$ M$_{\odot}$)  &    (km s$^{-1}$)  &  (km s$^{-1}$)  \\
\hline
LEDA126789                 &  2.339 $\pm$ 0.021  &  12.28 $\pm$ --     &    79.96 $\pm$ 0.19  & 30.76 $\pm$ 1.10 \\
SDSSJ125950.18-275445.4    &  2.133 $\pm$ 0.019  &  8.492 $\pm$ --     &    57.47 $\pm$ 0.20  &  37.2 $\pm$ 1.69 \\
SDSSJ130007.12-275551.4    &  1.476 $\pm$ 0.013  &  25.65 $\pm$ --     &    73.16 $\pm$ 0.52  & 48.86 $\pm$ 1.48 \\
SDSSJ125926.45-275124.7    &  1.346 $\pm$ 0.012  &  2.812 $\pm$ 0.259  &    58.06 $\pm$ 0.29  & 20.71 $\pm$ 2.35 \\
SDSSJ130026.16-280032.0    &   1.12 $\pm$ 0.010  &  8.492 $\pm$ --     &    47.36 $\pm$ 0.29  & 40.03 $\pm$ 1.40 \\
SDSSJ125914.43-280217.3    &  0.162 $\pm$ 0.006  &  3.707 $\pm$ --     &     16.2 $\pm$ 0.43  &  21.1 $\pm$ 1.42 \\
SDSSJ125953.93-275813.7    &  1.476 $\pm$ 0.013  &  9.312 $\pm$ --     &     55.6 $\pm$ 0.17  &  32.9 $\pm$ 1.25 \\
SDSSJ125636.78-271247.8    &  1.618 $\pm$ 0.028  &  3.707 $\pm$ --     &    40.25 $\pm$ 0.60  & 28.97 $\pm$ 3.05 \\
SDSSJ130000.97-275929.5    &  1.021 $\pm$ 0.009  &  21.33 $\pm$ --     &    47.33 $\pm$ 0.30  &  54.4 $\pm$ 1.40 \\
SDSSJ125844.58-274458.2    &  1.476 $\pm$ 0.013  &  17.74 $\pm$ --     &    44.28 $\pm$ 0.10  & 39.71 $\pm$ 1.15 \\
SDSSJ130042.86-280313.8    &  0.849 $\pm$ 0.007  &  2.339 $\pm$ --     &    34.21 $\pm$ 0.19  & 23.51 $\pm$ 2.84 \\
COMAi125949.960p275433     &  0.588 $\pm$ 0.005  &  10.21 $\pm$ --     &    43.51 $\pm$ 0.30  & 37.28 $\pm$ 1.42 \\
SDSSJ130032.61-280331.4    &  0.706 $\pm$ 0.012  &  11.2  $\pm$ --     &     30.2 $\pm$ 0.52  & 39.41 $\pm$ 1.27 \\
SDSSJ130036.58-275552.2    &  0.588 $\pm$ 0.005  &  4.887 $\pm$ --     &    36.76 $\pm$ 0.31  & 31.79 $\pm$ 2.14 \\
COMAi125713.240p272437     &  0.588 $\pm$ 0.005  &  7.064 $\pm$ --     &    37.44 $\pm$ 0.18  & 30.16 $\pm$ 1.26 \\
SDSSJ125942.36-280158.5    &  0.644 $\pm$ 0.011  &  5.358 $\pm$ --     &    31.93 $\pm$ 0.42  & 20.43 $\pm$ 0.96 \\
SDSSJ130027.57-280323.9    &  0.536 $\pm$ 0.009  &  1.618 $\pm$ --     &    25.91 $\pm$ 0.40  & 17.55 $\pm$ 1.86 \\
SDSSJ130004.03-280030.7    &  0.489 $\pm$ 0.004  &  7.064 $\pm$ --     &    36.71 $\pm$ 0.43  &  33.8 $\pm$ 1.53 \\
SDSSJ125902.43-280021.3    &  0.407 $\pm$ 0.004  &  5.875 $\pm$ --     &    36.47 $\pm$ 0.27  & 29.62 $\pm$ 1.64 \\
SDSSJ130018.70-275512.6    &  0.371 $\pm$ 0.003  &  4.065 $\pm$ --     &    31.06 $\pm$ 0.22  & 29.42 $\pm$ 1.45 \\
SDSSJ130044.10-280215.4    &  0.536 $\pm$ 0.009  &  5.875 $\pm$ --     &    25.56 $\pm$ 0.47  & 28.55 $\pm$ 1.10 \\
SDSSJ125943.53-275620.6    &  0.371 $\pm$ 0.010  &  9.312 $\pm$ --     &    26.72 $\pm$ 0.54  & 42.67 $\pm$ 1.55 \\
COMAi125828.358p271315     &  0.489 $\pm$ 0.009  &  3.381 $\pm$ --     &    27.16 $\pm$ 0.33  & 24.12 $\pm$ 1.35 \\
SDSSJ130042.51-280325.4    &  0.338 $\pm$ 0.006  &  7.745 $\pm$ --     &    32.48 $\pm$ 0.53  & 27.71 $\pm$ 0.89 \\
SDSSJ130037.30-275441.0    &  0.338 $\pm$ 0.009  &  8.492 $\pm$ --     &     27.2 $\pm$ 0.68  & 38.22 $\pm$ 1.53 \\
SDSSJ125955.93-275748.6    &  0.281 $\pm$ 0.002  &  2.565 $\pm$ --     &    25.61 $\pm$ 0.30  & 16.91 $\pm$ 1.31 \\
SDSSJ130032.96-275406.6    &  0.536 $\pm$ 0.009  &  11.2  $\pm$ --     &    21.68 $\pm$ 0.42  & 39.41 $\pm$ 1.27 \\
SDSSJ130003.18-275648.3    &  0.148 $\pm$ 0.003  &  1.476 $\pm$ 0.413  &    18.94 $\pm$ 0.28  & 15.75 $\pm$ 2.95 \\
SDSSJ125927.22-275257.0    &  0.195 $\pm$ 0.003  &  5.875 $\pm$ --     &    21.92 $\pm$ 0.30  & 18.71 $\pm$ 0.62 \\
SDSSJ125951.46-275935.4    &  0.213 $\pm$ 0.004  &  3.707 $\pm$ --     &    23.99 $\pm$ 0.33  & 24.04 $\pm$ 1.22 \\
SDSSJ125930.83-275810.2    &  0.234 $\pm$ 0.014  &  7.745 $\pm$ --     &    23.96 $\pm$ 1.30  & 35.52 $\pm$ 1.35 \\
SDSSJ125945.55-280313.4    &  0.162 $\pm$ 0.007  &  3.381 $\pm$ --     &    17.93 $\pm$ 0.56  & 17.78 $\pm$ 1.31 \\
COMAi13035.990p275505.4    &  0.162 $\pm$ 0.003  &  2.565 $\pm$ --     &    18.36 $\pm$ 0.31  & 22.19 $\pm$ 2.18 \\
SDSSJ130024.85-27.99085    &  0.162 $\pm$ 0.004  &  6.442 $\pm$ --     &    20.25 $\pm$ 0.46  & 31.02 $\pm$ 1.29 \\
SDSSJ125853.08-274741.8    &  0.234 $\pm$ 0.008  &  3.083 $\pm$ --     &    22.85 $\pm$ 0.61  & 22.94 $\pm$ 1.46 \\
SDSSJ125708.35-272923.9    &  0.148 $\pm$ 0.005  &  5.875 $\pm$ --     &    22.63 $\pm$ 0.75  & 31.06 $\pm$ 1.42 \\
SDSSJ125959.08-275841.4    &  0.281 $\pm$ 0.007  &  11.2  $\pm$ --     &    18.53 $\pm$ 0.62  & 43.89 $\pm$ 1.40 \\
SDSSJ130030.94-280312.8    &  0.085 $\pm$ 0.004  &  3.707 $\pm$ --     &    14.98 $\pm$ 0.65  &  24.3 $\pm$ 1.44 \\
SDSSJ130030.94-280312.8    &  0.093 $\pm$ 0.005  &  3.707 $\pm$ --     &    14.86 $\pm$ 0.76  & 23.13 $\pm$ 1.24 \\
COMAi125937.351p28210.6    &  0.085 $\pm$ 0.003  &  1.618 $\pm$ 0.149  &     17.2 $\pm$ 0.59  & 14.02 $\pm$ 1.66 \\
SDSSJ130034.32-275817.6    &  0.093 $\pm$ 0.002  &  1.476 $\pm$ 0.136  &    16.91 $\pm$ 0.30  &   15. $\pm$ 1.81 \\
SDSSJ125934.39-275942.9    &  0.148 $\pm$ 0.004  &  1.945 $\pm$ --     &    19.09 $\pm$ 0.46  & 16.17 $\pm$ 1.31 \\
SDSSJ130039.32-275748.4    &  0.093 $\pm$ 0.003  &  2.133 $\pm$ --     &     14.9 $\pm$ 0.43  & 16.88 $\pm$ 1.15 \\
COMAi13039.554p275350.0    &  0.123 $\pm$ 0.003  &  2.565 $\pm$ --     &    17.57 $\pm$ 0.49  & 22.92 $\pm$ 2.18 \\
SDSSJ125951.81-275726.3    &  0.135 $\pm$ 0.012  &  5.875 $\pm$ --     &    15.34 $\pm$ 1.44  & 31.18 $\pm$ 1.68 \\
COMAi125925.477p28211.0    &  0.071 $\pm$ 0.006  &  1.476 $\pm$ 0.136  &    13.8  $\pm$ 0.97  & 11.72 $\pm$ 2.03 \\
SDSSJ125700.89-273155.1    &  0.102 $\pm$ 0.011  &  2.339 $\pm$ --     &    11.54 $\pm$ 0.88  & 21.01 $\pm$ 1.47 \\
SDSSJ130030.94-280312.8    &  0.085 $\pm$ 0.004  &  3.707 $\pm$ --     &    14.98 $\pm$ 0.65  &  24.3 $\pm$ 1.44 \\
SDSSJ130030.94-280312.8    &  0.093 $\pm$ 0.005  &  3.707 $\pm$ --     &    14.86 $\pm$ 0.76  & 23.13 $\pm$ 1.24 \\
SDSSJ130029.81-280401.0    &  0.085 $\pm$ 0.002  &  1.021 $\pm$ 0.094  &    14.54 $\pm$ 0.31  & 12.18 $\pm$ 1.81 \\
SDSSJ130004.04-275342.7    &  0.281 $\pm$ 0.015  &  5.875 $\pm$ --     &    14.38 $\pm$ 0.72  & 30.94 $\pm$ 1.41 \\
SDSSJ125952.18-275946.3    &  0.078 $\pm$ 0.002  &  1.228 $\pm$ 0.113  &    15.05 $\pm$ 0.40  & 15.15 $\pm$ 2.24 \\
SDSSJ125856.78-274644.5    &  0.049 $\pm$ 0.001  &  0.234 $\pm$ 0.088  &    13.76 $\pm$ 0.31  &  3.34 $\pm$ 2.08 \\
\hline                             
\end{tabular}                      
\label{table:table4}               
\end{table*}

\section*{Acknowledgments}

We would like to kindly thank Jeremiah Ostriker, Alister Graham and Mordecai Mark Mac-Low 
for useful discussions and suggestions.  NL acknowledges support from an NSERC Postdoctoral Fellowship.


\bsp

\label{lastpage}


\begin{thebibliography}{99}

\bibitem[\protect\citeauthoryear{Antonini et al.}{2012}]{antonini12} Antonini F., Capuzzo-Dolcetta R., 
Mastrobuono-Battisti A., Merritt D. 2012, ApJ, 750, 111
\bibitem[\protect\citeauthoryear{Antonini}{2013}]{antonini13} Antonini F. 2013, 
ApJ, 763, 62
\bibitem[\protect\citeauthoryear{Beasley et al.}{2009}]{beasley09} Beasley M. A., Cenarro A. J., 
Strader J., Brodie J. P. 2009, AJ, 137, 5146
\bibitem[\protect\citeauthoryear{Bell et al.}{2003}]{bell03} Bell E. F.,
McIntosh D. H., Katz N., Weinberg M. D. 2003, ApJS, 149, 289
\bibitem[\protect\citeauthoryear{Binggeli, Sandage \& Tarenghi}{1984}]{binggeli84} Binggeli B., 
Sandage A., Tarenghi M. 1984, AJ, 89, 64
\bibitem[\protect\citeauthoryear{Binney \& Tremaine}{1987}]{binney87}
  Binney J., Tremaine S., 1987, Galactic Dynamics (Princeton:
  Princeton University Press)
\bibitem[\protect\citeauthoryear{Balcells}{2007}]{balcells07} Balcells M., Graham A. W., 
Peletier R. F. 2007, ApJ, 665, 1084
\bibitem[\protect\citeauthoryear{Blakeslee et al.}{2002}]{blakeslee02} Blakeslee J. P., 
Lucey J. R., Tonry J. L., Hudson M. J., Narayan V. K., Barris B. J. 2002, MNRAS, 330, 443
\bibitem[\protect\citeauthoryear{Bruzual \& Charlot}{2003}]{bruzual03} Bruzual G., 
Charlot S. 2003, MNRAS, 344, 1000
\bibitem[\protect\citeauthoryear{Busarello, Longo \& Feoli}{1992}]{busarello92} Busarello G., 
Longo G., Feoli A. 1992, A\&A, 262, 52
\bibitem[\protect\citeauthoryear{Cappellari et al.}{2006}]{cappellari06}
Cappellari M. et al.  2006, MNRAS, 366, 1126
\bibitem[\protect\citeauthoryear{Cappellari et al.}{2013a}]{cappellari13a} Cappellari M., 
Scott N., Alatalo K., Blitz L., Bois M., Bournaud F., Bureau M., Crocker A. F., 
Davies R. L., Davis T. A., de Zeeuw P. T., Duc P.-A., Emsellem E., Khochfar S., 
Krajnovic D., Kuntschner H., McDermind R. M., Morganti R., Naab T., Oosterloo T., 
Sarzi M., Serra P., Weijmans A.-M., Young L. M. 2013a, MNRAS, 432, 1709
\bibitem[\protect\citeauthoryear{Cappellari et al.}{2013b}]{cappellari13b} Cappellari M.,
McDermind R. M., Alatalo K., Blitz L., Bois M., Bournaud F., Bureau M., Crocker A. F.,
Davies R. L., Davis T. A., de Zeeuw P. T., Duc P.-A., Emsellem E., Khochfar S.,
Krajnovic D., Kuntschner H., Morganti R., Naab T., Oosterloo T.,
Sarzi M., Scott N., Serra P., Weijmans A.-M., Young L. M. 2013b, MNRAS, 432, 1862
\bibitem[\protect\citeauthoryear{Capuzzo-Dolcetta \& Miocchi}{2008}]{capuzzo-dolcetta08}
Capuzzo-Dolcetta R., Miocchi P. 2008, MNRAS, 388, 69
\bibitem[\protect\citeauthoryear{Carter et al.}{2008}]{carter08} Carter D., et al. 2008, 
ApJS, 176, 424
\bibitem[\protect\citeauthoryear{Chae, Bernardi \& Kravstov}{2014}]{chae14} Chae K.-H., 
Bernardi M., Kravstov A. V. 2014, MNRAS, in press (arXiv:1305.5471) 
\bibitem[\protect\citeauthoryear{Cote et al.}{2004}]{cote04} Cote P., Blakeslee J. P.,
Ferrarese L., Jordan A., Mei S., Merritt D., Milosavljevic M., Peng E. W., Tonry J. L.,
West M. J. 2004, ApJS, 153, 223
\bibitem[\protect\citeauthoryear{Cote et al.}{2006}]{cote06} Cote P., Piatek S.,
Ferrarese L., Jordan A., Merritt D., Peng E. W., Hasegan M., Blakeslee J. P.,
Mei S., West M. J., Milosavljevic M., Tonry J. L. 2006, ApJS, 165, 57
\bibitem[\protect\citeauthoryear{den Brok et al.}{2014}]{denbrok14} den Brok M., et al. 
2014, MNRAS, 445, 2385
\bibitem[\protect\citeauthoryear{Emsellem et al.}{2007}]{emsellem07} Emsellem E., 
Cappellari M., Krajnovic D., van de Ven G., Bacon R., Bureau M., Davies R. L., de Zeeuw P. T., 
Falcon-Barroso J., Kuntschner H. 2007, MNRAS, 379, 401
\bibitem[\protect\citeauthoryear{Gnedin, Ostriker \& Tremaine}{2014}]{gnedin14} Gnedin O. Y., 
Ostriker J. P., Tremaine S. 2014, ApJ, 785, 71
\bibitem[\protect\citeauthoryear{Graham et al.}{2005}]{graham05} Graham A. W., Driver S. P., 
Petrosian V., Conselice C. J., Bershady M. A., Crawford S. M., Goto T. 2005, AJ, 130, 1535
\bibitem[\protect\citeauthoryear{Ferrarese et al.}{2006}]{ferrarese06b}
Ferrarese L., Cote P., Jordan A., Peng E. W., Blakeslee J. P., Piatek S., Mei S.,
Merritt D., Milosavljevic M., Tonry J. L., West M. J. 2006, ApJS, 164, 334
\bibitem[\protect\citeauthoryear{Forbes et al.}{2008}]{forbes08} Forbes D. A., 
Lasky P., Graham A. W., Spitler L. 2008, MNRAS, 389, 1924 
\bibitem[\protect\citeauthoryear{Georgiev \& B\"oker}{2014}]{georgiev14} Georgiev I. Y., 
B\"oker T. 2014, arXiv:1404.5956
\bibitem[\protect\citeauthoryear{Graham \& Worley}{2008}]{graham08} Graham A. W.,
Worley. C. C. 2008, MNRAS, 388, 1708
\bibitem[\protect\citeauthoryear{Graham \& Spitler}{2009}]{graham09} Graham A. W.,
Spitler L. R. 2009, MNRAS, 397, 2148
\bibitem[\protect\citeauthoryear{Graham}{2012}]{graham12} Graham A. W. 2012, 
MNRAS, 422, 1586
\bibitem[\protect\citeauthoryear{Harris}{1996, 2010 update}]{harris96}
  Harris, W. E. 1996, AJ, 112, 1487 (2010 update)
\bibitem[\protect\citeauthoryear{Hartmann et al.}{2011}]{hartmann11} Hartmann M., 
Debattista V. P., Seth A., Cappellari M., Quinn T. R. 2011, MNRAS, 418, 2697
\bibitem[\protect\citeauthoryear{Jaffe}{1983}]{jaffe83} Jaffe W. 1983, MNRAS, 202, 995
\bibitem[\protect\citeauthoryear{Jeffreys}{1946}]{jeffreys46} Jeffreys H. 1946, RSPSA, 186, 453 
\bibitem[\protect\citeauthoryear{Jordan et al.}{2007}]{jordan07} Jordan A., McLaughlin D. E.,
Cote P., Ferrarese L., Peng E. W., Mei S., Villegas D., Merritt D., Tonry J. L.,
West M. J. 2007, ApJS, 171, 101
\bibitem[\protect\citeauthoryear{Kormendy \& Kennicutt}{2004}]{kormendy04} Kormendy J.,
Kennicutt R. C. Jr. 2004, ARA\&A, 42, 603
\bibitem[\protect\citeauthoryear{Kroupa}{2001}]{kroupa01} Kroupa P. 2001, MNRAS, 322, 231
\bibitem[\protect\citeauthoryear{Laurikainen et al.}{2011}]{laurikainen11} Laurikainen E., 
Salo H., Buta R., Knapen J. H. 2011, MNRAS, 418, 1452
\bibitem[\protect\citeauthoryear{Leigh, B\"oker \& Knigge}{2012}]{leigh12} Leigh N. W. C., 
B\"oker T., Knigge C. 2012, MNRAS, 424, 2130
\bibitem[\protect\citeauthoryear{Long, Charles \& Dubus}{2002}]{long02} Long K. S.,
Charles P. A., Dubus G., 2002, ApJ, 569, 204
\bibitem[\protect\citeauthoryear{McLaughlin, King \& Nayakshin}{2006}]{mclaughlin06}
McLaughlin D. E., King A. R., Nayakshin S. 2006, ApJL, 650, L37
\bibitem[\protect\citeauthoryear{Merritt}{2013}]{merritt13} Merritt D. 2013, Dynamics and Evolution
of Galactic Nuclei (Princeton:  Princeton University Press)
\bibitem[\protect\citeauthoryear{Miller \& Davies}{2012}]{miller12} Miller M. C., 
Davies M. B. 2012, ApJ, 755, 81
\bibitem[\protect\citeauthoryear{Nayakshin, Wilkinson \& King}{2009}]{nayakshin09}
Nayakshin S., Wilkinson M. I., King A. 2009, MNRAS, 398, 54
\bibitem[\protect\citeauthoryear{Nguyen et al.}{2014}]{nguyen14} Nguyen D. D., Seth A. C., 
Reines A. E., den Brok M., Sand D., McLeod B. 2014, ApJ, 794, 34
\bibitem[\protect\citeauthoryear{Pota et al.}{2013}]{pota13} Pota V., Forbes D. A., 
Romanowsky A. J., Brodie J. P., Spitler L. R., Strader J., Foster C., Arnold J. A., 
Benson A., Blom C., Hargis J. R., Rhode K. L., Usher C. 2013, MNRAS, 428, 389 
\bibitem[\protect\citeauthoryear{Quinlan \& Shapiro}{1990}]{quinlan90} Quinlan G. D.,
Shapiro S. L. 1990, ApJ, 356, 483
\bibitem[\protect\citeauthoryear{Scott \& Graham}{2013}]{scott13} Scott N., 
Graham A. W. 2013, ApJ, 763, 76
\bibitem[\protect\citeauthoryear{Scott et al.}{2014}]{scott14} Scott N., Davies R. L., 
Houghton R. C. W., Cappellari M., Graham A. W., Pimbblet K. A. 2014, MNRAS, 441, 274
\bibitem[\protect\citeauthoryear{Seth et al.}{2008}]{seth08} Seth A., Agueros M.,
Lee D., Basu-Zych A. 2008, ApJ, 678, 116
\bibitem[\protect\citeauthoryear{Spitzer}{1969}]{spitzer69} Spitzer L. Jr. 1969, 
ApJL, 158, L139
\bibitem[\protect\citeauthoryear{Toloba et al.}{2014a}]{toloba14a} Toloba E., 
Guhathakurta P., Peletier R., Boselli A., Lisker T., Falcon-Barroso J., Simon J., 
van de Ven G., Paudel S., Emsellem E., Janz J., den Brok M., Gorgas J., 
Hensler G., Laurikainen E., Niemi S. M., Rys A., Salo H. 2014, ApJS, 
accepted (arXiv:1410.1550)
\bibitem[\protect\citeauthoryear{Toloba et al.}{2014b}]{toloba14b} Toloba E.,
Guhathakurta P., Boselli A., Peletier R., Emsellem E., Lisker T., van de Ven G., 
Simon J., Falcon-Barroso J., Adams J., Benson A., Boissier S., den Brok M., 
Gorgas J., Hensler G., Janz J., Laurikainen E., Paudel S., Rys A., Salo H. 2014, ApJ,
submitted (arXiv:1410.1552)
\bibitem[\protect\citeauthoryear{Tonry et al.}{2001}]{tonry01} Tonry J. L., Dressler A., 
Blakeslee J. P., Ajhar E. A., Fletcher A. B., Luppino G. A., Metzger M. R. 2001, ApJ, 546, 681
\bibitem[\protect\citeauthoryear{Tremaine, Ostriker \& Spitzer}{1975}]{tremaine75}
Tremaine S. D., Ostriker J. P., Spitzer L. Jr. 1975, ApJ, 196, 407
\bibitem[\protect\citeauthoryear{Watson et al.}{2005}]{watson05} Watson L. C., 
Martini P., Lisenfeld U., Wong M.-H., B\"oker T., Schinnerer E. 2005, ApJ, 751, 123 

\end{thebibliography}
\end{document}